\documentclass[12pt]{article}
\pdfoutput=1
\usepackage{putex}
\usepackage{amsmath}
\usepackage{amssymb}
\usepackage{graphicx}
\usepackage{epstopdf}
\usepackage{enumerate}
\usepackage{cite}
\usepackage{tensor}
\usepackage{slashed}
\usepackage{feynmf}
\usepackage{hyperref}

\usepackage[utf8x]{inputenc}
\usepackage{braket}
\usepackage{titlesec}
\usepackage{fixltx2e}
\usepackage{cleveref}
\hypersetup{breaklinks}

\newcommand{\floor}[1]{\lfloor #1 \rfloor}

\usepackage[usenames,dvipsnames,svgnames,table]{xcolor}
\def\GcRb{\overline{\Gc_R}}
\def\hsl{hs[$\lambda$]}

\newcommand{\e}[2] {\begin{equation} \label{#1} #2 \end{equation}}
\newcommand{\es}[2] {\begin{equation} \label{#1} \begin{split} #2 \end{split} \end{equation}}
\def\eqr{\eqref}

\def\subsec{\subsection}
\def\subsubsec{\subsubsection}
\def\asym{{\rm \bf asym}}
\def\Oc{{\cal O}}
\def\l{\lambda}
\def\vs{\vskip .1 in}
\def\hw{\text{hw}}
\def\a{\alpha}
\def\L{\Lambda}
\def\rar{\rightarrow}
\def\rhor{|\rho\rangle}

\def\Ab{\overline{A}}
\def\ab{\overline{a}}
\newcommand {\be} {\begin {equation}}
\newcommand {\ee} {\end {equation}}
\newcommand{\bea}{\begin{eqnarray}}
\newcommand{\eea}{\end{eqnarray}}
\def\Tr{{\rm{Tr}}}
\def\eps{\epsilon}

\def\wb{\overline{w}}
\def\rt{\rightarrow}

\def\Gc{{\cal G}}

 \newmuskip\pFqmuskip

\newcommand*\pFq[6][8]{%
  \begingroup 
  \pFqmuskip=#1mu\relax
  \mathcode`\,=\string"8000
  \begingroup\lccode`\~=`\,
  \lowercase{\endgroup\let~}\pFqcomma
  {}_{#2}F_{#3}{\left[\genfrac..{0pt}{}{#4}{#5};#6\right]}%
  \endgroup
}
\newcommand{\pFqcomma}{\mskip\pFqmuskip}

\makeatletter
\renewcommand{\@maketitle}{
\newpage
 \begin{center}%
  {\large\bfseries \@title \par}%
 \end{center}%
 \par} \makeatother

\numberwithin{equation}{section}

\titleformat*{\section}{\large\bfseries}

\begin{document}

\institution{UCLA}{Department of Physics and Astronomy, University of California, Los Angeles, CA 90095, USA}

\institution{PU}{Department of Physics, Princeton University, Princeton, NJ 08544, USA}

\title{General Results for Higher Spin Wilson Lines \\and Entanglement in Vasiliev Theory}

\authors{Ashwin Hegde\worksat{\UCLA}, Per Kraus\worksat{\UCLA}, Eric Perlmutter\worksat{\PU}}

\abstract{We develop tools for the efficient evaluation of Wilson lines in 3D higher spin gravity, and use these to compute entanglement entropy in the hs$[\lambda]$ Vasiliev theory that governs the bulk side of the duality proposal of Gaberdiel and Gopakumar.   Our main technical advance is the determination of SL(N) Wilson lines for arbitrary $N$, which, in suitable cases, enables us to analytically continue to hs$[\lambda]$ via $N \rightarrow -\lambda$. We apply this result to compute various quantities of interest, including entanglement entropy expanded perturbatively in the background higher spin charge, chemical potential, and interval size. This includes a computation of entanglement entropy in the higher spin black hole of the Vasiliev theory. These results are consistent with conformal field theory calculations. We also provide an alternative derivation of the Wilson line, by showing how it arises naturally from earlier work on scalar correlators in higher spin theory. The general picture that emerges is consistent with the statement that the SL(N) Wilson line computes the semiclassical $W_N$  vacuum block, and our results provide an explicit result for this object.}

\date{}

\maketitle
\setcounter{tocdepth}{2}
\tableofcontents
\section{Introduction}
Entanglement, it has been suggested, builds spacetime \cite{Ryu:2006bv, Swingle:2009bg, VanRaamsdonk:2009ar, VanRaamsdonk:2010pw, Czech:2012bh, Hubeny:2012wa, Maldacena:2013xja,Balasubramanian:2013lsa,
Faulkner:2013ica,Balasubramanian:2014sra, Headrick:2014cta, Almheiri:2014lwa, Czech:2015qta}.  We now have a robust holographic entanglement dictionary in the context of Einstein gravity and its higher derivative generalizations. The Ryu-Takayanagi formula \cite{Ryu:2006bv} and its relatives \cite{Hubeny:2007xt, Lewkowycz:2013nqa, Dong:2013qoa,  Camps:2013zua, Castro:2014tta} encode the quantum information of CFT states in bulk geometry; conversely, inequalities obeyed by entanglement entropy in holographic CFTs may be transmuted into dynamical gravitational laws, defining consistent dynamics in weakly curved AdS spacetime \cite{
Faulkner:2013ica, Swingle:2014uza, Lin:2014hva, Lashkari:2014kda, Bhattacharya:2014vja, Lashkari:2015hha}.

The slogan that entanglement builds spacetime has yet to elucidate what becomes of smooth spacetime geometry far away from the classical Einstein gravity regime. In a UV complete theory of gravity in AdS, such as string or M-theory, new physics at the string or Planck scale ensures that classical spacetime concepts, such as the Ryu-Takayanagi formula, cease to be meaningful. One might hope that entanglement remains a good observable for constructing quantum geometry.  The AdS/CFT correspondence suggests that CFT entanglement may be used to define these concepts from the boundary inwards: as a CFT may be regarded as the non-perturbative definition of a quantum string theory in AdS $\times~ {\cal M}$, perhaps CFT entanglement may be likewise regarded as defining what we mean by ``spacetime'' in that theory.

A highly ambitious goal, but perhaps a reasonable place to start, is to find the generalization of Ryu-Takayanagi to string theory at finite $\ell_s$. In string theory, there is, at energies of order $1/\ell_s$, no hierarchy between the graviton and higher spin modes of a closed string. As the infinite towers of higher spin modes become massless, the theory is believed to acquire a huge symmetry enhancement \cite{Sundborg:2000wp,Bianchi:2003wx,Sagnotti:2003qa,Sagnotti:2013bha}. In type IIB string theory on AdS$_3 \times S^3 \times T^4$, for instance, we now have some of our first data on what this symmetry algebra may be \cite{Gaberdiel:2014cha,Gaberdiel:2015mra,Baggio:2015jxa,Gaberdiel:2015uca}; this raises the tantalizing prospect of reformulating string theory on AdS$_3 \times S^3 \times T^4$ as a higher spin theory, highly (or even uniquely) constrained by its symmetries. In this setting, it is not at all clear what becomes of spacetime, much less what role entanglement plays and how to compute it.

This discussion suggests that if we want to compute entanglement in bona fide AdS$_3$ string theory, studying entanglement in AdS$_3$ higher spin theories is a good start. At present, the Vasiliev theories of higher spin gravity are the only explicitly constructed examples of higher spin gauge fields coupled to matter \cite{Prokushkin:1998bq,Bekaert:2005vh,Didenko:2014dwa}. Morally speaking (and literally so in the case of AdS$_3 \times S^3 \times T^4$ \cite{Gaberdiel:2014cha,Gaberdiel:2015mra,Baggio:2015jxa}), the Vasiliev degrees of freedom model the lowest of the many infinite towers of higher spin fields of string theory. The simplest non-supersymmetric Vasiliev theory contains one such tower. The dynamics of these higher spin gauge fields are governed by two copies of the higher spin algebra \hsl, which acts as the toy model for the stringy symmetries \cite{Prokushkin:1998bq}. Just as in high energy string theory, it is not known what the proper gauge-invariant generalization of spacetime geometry is in the presence of the Vasiliev gauge symmetries \cite{Vasiliev:2003ar,Vasiliev:2001dc,Ammon:2011nk}. With the above as inspiration, we will, in this paper, take the modest step of performing the first computations of entanglement entropy in Vasiliev's theory of higher spin gravity in AdS$_3$.

Lofty inspiration aside, the computation of entanglement in 3D Vasiliev theory has been sought from other, holographic, vantage points. An even simpler theory of 3D higher spins, without matter, is given by an  SL(N,$\mathbb{R}$) $\times$ SL(N,$\mathbb{R}$) Chern-Simons theory \cite{Blencowe:1988gj,Blencowe:1988gj,Campoleoni:2011hg}, generalizing the usual construction of gravity as an SL(2,$\mathbb{R}$) $\times$ SL(2,$\mathbb{R}$) Chern-Simons theory \cite{Achucarro:1987vz, Witten:1988hc}. In AdS, these theories give rise to (two) asymptotic $W_N$ algebras, and hence they capture the large $c$ dynamics of $W_N$ currents, independent of any particular CFT realization. In \cite{Ammon:2013hba,deBoer:2013vca}, two entanglement functionals were proposed in the SL(N) theories, and later proven to be equivalent \cite{Castro:2014mza}. These purport to compute the single interval entanglement entropy of a putative dual CFT with $W_N$ symmetry; there is now plenty of evidence for this \cite{Datta:2014ska,Datta:2014uxa,Long:2014oxa,Datta:2014zpa,deBoer:2014sna}. An outstanding goal in the field has been to generalize this to the infinite-dimensional \hsl$\times$\hsl\ Chern-Simons theory that generates the current sector of the Vasiliev theory.

To set up our Vasiliev calculations, we need to recall what has been done for SL(N). As we review in more detail in Section \ref{sec2}, the SL(N) entanglement prescription is to compute a certain bulk Wilson line for the two Chern-Simons connections, anchored on the asymptotic boundary at the endpoints of the entanglement interval. Defining the Wilson line for the SL(N)$\times $ SL(N) connection ${\cal A}$, 
\be\label{wilson1int}
W_{\cal R}(C)=\Tr_{\cal R}\left( {\cal P} \exp \int_C {\cal A}  \right)~,
\ee
the claim is that $S_{EE} = -\log (W_{\cal R}(C))$ for a suitably chosen representation ${\cal R}$.\footnote{We will not work directly with this Wilson line expression --- and hence will not define it precisely --- but rather a different object than can be argued \cite{Ammon:2013hba,deBoer:2013vca} to represent the Wilson line in the large $c$ limit.}  This representation has quantum numbers that grow with large $c$, hence allowing us to apply a semiclassical approximation to $W_{\cal R}(C)$. A natural question, answered in \cite{Hijano:2014sqa,deBoer:2014sna}, is what this computes when the representation is modified. The answer proposed in \cite{Hijano:2014sqa,deBoer:2014sna} is simple: the Wilson line computes a bulk two-point function for an operator carrying the higher spin charges needed to specify the representation ${\cal R}$, in the background specified by the connection ${\cal A}$. 
Furthermore, this correlation function is equivalent to the $W_N$ four-point conformal block for the identity module in a certain large $c$ limit, given in \eqr{sclim}.

The fact that so much dynamical information about SL(N) higher spin gravity can be ascertained from symmetry considerations has recent precedent in ordinary 3D gravity. \cite{Fitzpatrick:2014vua,Hijano:2015rla,Fitzpatrick:2015zha,Hijano:2015qja} studied probe scalar two-point functions in locally AdS$_3$ geometries. Thinking of the background as generated by a heavy operator $\Oc_H$, and the probe scalar as dual to some ``light'' operator $\Oc_L$, the leading order exchange of gravitons between the probe and the surrounding geometry captures the exchange of the Virasoro vacuum module in a four-point function at large $c$. By independent CFT calculations of the four-point Virasoro vacuum block, ${\cal F}_{\rm vac}$, \cite{Fitzpatrick:2014vua} demonstrated this mapping. Schematically, this universality can be written as
\e{univ}{e^{-S_{\rm geodesic}} ~~ \approx ~~ \langle \Oc_H|\Oc_L\Oc_L|\Oc_H\rangle ~~ \approx ~~ |{\cal F}_{\rm vac}|^2}
where ${\cal F}_{\rm vac}$ is evaluated in a large $c$ limit
\e{}{c\rar\infty~, \quad {h_H\over c}, {h_L\over c}~\text{fixed}~,\quad{h_L\over c}\ll 1~.}
The SL(N) version of this, described above, just upgrades everything to include higher spin charges, and replaces the probe worldline action by the Wilson line:
\e{univ2}{W_{\cal R}(C) ~~ \approx ~~ \langle \Oc_H|\Oc_L\Oc_L|\Oc_H\rangle ~~ \approx ~~ |{\cal F}_{\rm vac}^{\,W_N}|^2}
(The precise large $c$ limit is now given by \eqr{sclim}.) The vacuum dominance of semiclassical correlation functions is believed to hold for every sparse, large $c$ CFT, assuming that all operator dimensions scale with $c$ \cite{Hartman:2013mia}.

In this paper we will firm up these notions of universality in higher spin gravity, on our way to computing entanglement entropy in Vasiliev theory. To set the stage, let us highlight some outstanding problems for higher spin Wilson lines, all of which we will tackle in turn.

The first is that the SL(N) Wilson line, for arbitrary representations ${\cal R}$ and in arbitrary higher spin backgrounds, has not actually been explicitly computed for arbitrary $N$. Calculations in \cite{deBoer:2014sna} only treat the $N=3$ case in detail. Equivalently, the determination of the semiclassical $W_N$ vacuum block remains an open problem.

A second open problem is that the SL(N) Wilson line has never been derived from the field equations of a bona fide SL(N) theory coupled to matter.  Rather,  it has  been motivated by the role of Wilson lines as gauge-covariant observables in Chern-Simons theory, and verified to produce sensible results.  It would be satisfying to understand its connection to first principles computations based on the field equations coupling matter to higher spins.

Finally, as we have emphasized, the extension to \hsl, and hence to the gauge sector of Vasiliev theory, has not been done. The \hsl\ Wilson line computes the semiclassical $W_{\infty}[\l]$ vacuum block, where $W_{\infty}[\l]$ is the asymptotic symmetry algebra of \hsl\ gravity in AdS$_3$ \cite{Henneaux:2010xg, Gaberdiel:2011wb}. It is difficult to directly evaluate the \hsl\ Wilson line because \hsl\ is an infinite-dimensional algebra, and we will not fully solve this problem here. However, as we describe below, in perturbation theory in the higher spin charges one can skirt these difficulties by using a certain analytic continuation from the SL(N) Wilson line, once the latter is known for general $N$.

Before proceeding to a summary of our results, let us note that another route towards establishing the link between the Wilson line and CFT results goes through Toda field theory.  Just as Liouville theory captures the universal information dictated by Virasoro symmetry, Toda theory does the same for $W_N$ symmetry \cite{deBoer:1991jc,Fateev:2007ab}.  \cite{deBoer:2014sna} showed how to compute the semiclassical $W_N$ vacuum block in Toda theory, with the answer being expressed in terms of determinants of SL(N) matrices.   This same result can be shown to arise from the Wilson line \cite{Castropc}. The determinants arise in the same way as in (\ref{detform}).

\subsec{Summary of results}
We now give an extended summary of our results, which can be found in Sections \ref{wilson line I}--\ref{wnsec}. These are preceded by a brief review in Section \ref{sec2} of 3D higher spin gravity and the circle of ideas relating higher spin Wilson lines, $W_N$ conformal blocks and universality of correlation functions at large $c$.

\subsubsec{The explicit SL(N) Wilson line (Sections \ref{wilson line I}--\ref{closed})}

We fully determine the bulk Wilson line in the SL(N) higher spin theory, for an arbitrary probe propagating in an asymptotically AdS background. This builds on work of \cite{deBoer:2014sna}, where only the $N=3$ case was explicitly computed. This result thereby gives the explicit semiclassical $W_N$ vacuum block, for arbitrary charges subject to the large $c$ limit \eqr{sclim}.

The computation begins in Section \ref{wilson line I} by showing that the usual expression for the Wilson line, presented in \cite{Ammon:2013hba,deBoer:2013vca}, can be drastically simplified in the near-boundary limit. Recall that the near-boundary limit is the physically relevant regime: in analogy to the use of the GKPW prescription to extract CFT correlators from AdS amplitudes, one must take the near-boundary limit of the Wilson line to extract a semiclassical CFT correlator. One of the primary difficulties in computing the Wilson line so far is that the eigenvalues of the connection depend in a complicated way on the radial coordinate, and the near-boundary limit may only be taken after computing the full bulk Wilson line. What we have derived is a direct expression for the near-boundary result.

This result can be summarized as follows. The Wilson line anchored at points $(0,w)$ on the boundary, where $w$ is a complex coordinate, is evaluated in a background specified by a pair of flat, constant SL(N) connections $(a,\ab)$. The probe sits in a representation $R$ of SL(N), which can be specified by a charge vector $q$.\footnote{Note that we are now denoting the representation as $R$.  This is because in (\ref{Iqfactorint}) we have pulled out a factor of $k_{CS}$ corresponding to rescaling $q$.  The weight vector $q$ will thus be thought of as being $O(1)$, whereas the highest weight vector for ${\cal R}$ would be $O(k_{CS})$.}    The Wilson line action $I_q = -\log W_q$ in the near-boundary limit, call it $[I_q]_{\rm finite}$, is then
\be\label{Iqfactorint}
[I_q]_{\rm finite} =k_{CS} \log  \Gc_R+k_{CS} \log \GcRb
\ee
where $\Gc_R$ and $\overline{\Gc_R}$ are the matrix elements
\e{gcrint}{\Gc_R =  \langle -{\rm hw}_R| e^{\L} | {\rm hw}_R\rangle~, \quad \overline{\Gc_R} = \langle {\rm hw}_R| e^{-\overline{\L}} \ket{-{\rm hw}_R}~.}
$(\L,\overline{\L})$ are linear combinations of the respective bulk connections,
\e{lambint}{\Lambda \equiv a_w w +a_{\wb}\wb~, \quad \overline{\Lambda} \equiv \ab_w w + \ab_{\wb}\wb~;}
$\ket{\hw_R}$ and $\ket{-\hw_R}$ are highest and lowest weight states, respectively, of $R$; and $k_{CS} \propto c$ is the Chern-Simons level. 

To make this even clearer, we may write it in terms of the charge vector $q$, which is parameterized by a set of $N-1$ real numbers, $d_i $.  When $d_i \in \mathbb{Z}^+$, the charge vector $q$ is the weight of a highest weight state $|{\rm hw}_R\rangle$ of a finite-dimensional representation, and the $d_i$ are the Dynkin labels.  In fact, we will want to deal with more general probes such that $d_i \in \mathbb{R}$.  These representations are generically infinite-dimensional, and so the formulas in (\ref{gcrint}) do not apply directly since there is no lowest weight state $\ket{-{\rm hw}_R}$. Our prescription is that we compute with general $d_i \in \mathbb{Z}^+$, and then continue to arbitrary  $d_i$ in the final result.  Using (\ref{wilbint}) below, this step is rather trivial to implement. The Dynkin labels are in one-to-one correspondence with  Young diagrams of SL(N). Just as a Young diagram is symmetrized among its columns, the matrix element $\Gc_R$ may be simply expressed in terms of the matrix elements of the $k$-box antisymmetric representations, $\Gc_k$:
\e{wilbint}{\Gc_R = \prod_{k=1}^{N-1}\Gc_k^{\,d_k}~, \quad \text{where}~~~\Gc_k \equiv \bra{-\hw}e^{\Lambda}|\hw\rangle_k}
An analogous formula holds for $\GcRb$. Altogether, \eqr{Iqfactorint} and \eqr{wilbint} are significantly simpler than previous prescriptions.

Moreover, we can explicitly evaluate \eqr{wilbint} for arbitrary representations. $\Gc_R$ may be expressed solely in terms of the eigenvalues of $\L$. In Section \ref{closed}, we evaluate $\Gc_R$ explicitly in terms of these eigenvalues and the charges $d_i$ of the probe. This, then, provides the final explicit expression for the Wilson line, and hence the semiclassical $W_N$ vacuum block, in terms of the light and heavy higher spin charges. The result can be found in equation \eqr{Gkresult}.

Note that the result \eqr{Iqfactorint} holomorphically factorizes. This reflects the fact that the semiclassical correlator $\langle \Oc_H|\Oc_L\Oc_L|\Oc_H\rangle$ is believed to be dominated by the semiclassical $W_N$ vacuum block. We can thus read off the block as
\e{}{{\cal F}_{\rm vac}^{\,W_N} = \Gc_{R}^{-k_{CS}}}

Because the result for the block applies in any large $c$ CFT with $W_N$ symmetry, this is a useful result beyond the holographic context. Examples of such theories include Toda theory at large $c$, or large $N$ symmetric product CFTs, Sym$^N(X)$, where the seed theory $X$ has $W_N$ symmetry.\footnote{The latter theory has a much larger chiral algebra, of which $W_N$ is its ``diagonal'' subalgebra.}

\subsubsec{``Deriving" SL(N) Wilson lines from SL(N) Vasiliev theory (Section \ref{vasilievwilsonlines})}

The lack of a derivation of the Wilson line, alluded to earlier, is partly due to the paucity of theories which actually feature SL(N) gauge fields consistently coupled to matter. In fact, we know of only one: the Vasiliev theory at $\l=\pm N$, with all fields of spin $s>N$ truncated. This is possible because hs[$\pm$N]$/\chi_N \cong$ SL(N), where the spin $s>N$ fields form an ideal, $\chi_N$. We call this theory ``SL(N) Vasiliev theory.''  In Section \ref{vasilievwilsonlines}, we provide arguments that motivate the appearance of the  Wilson line from the field equations of SL(N) Vasiliev theory. The argument is based on the results of \cite{Hijano:2013fja}. The key point is that $\Gc_R$ appearing in \eqr{gcrint} is precisely the formula for a two-point function in SL(N) Vasiliev theory. This fact was first derived in \cite{Hijano:2013fja} by direct expansion of the Vasiliev master field equations around asymptotically AdS higher spin backgrounds.  As we will discuss, the equivalence of the Wilson line and the correlator computed in \cite{Hijano:2013fja} is not automatic and requires some justification, as it involves an extrapolation of the charges carried by the objects from one regime to another.  

\subsubsec{Entanglement in \hsl\ Vasiliev theory (Sections \ref{smallcharge}--\ref{shortint})}

The SL(N) Wilson line can be applied to compute correlators/vacuum blocks for any set of charges. An especially interesting case is that of a probe with vanishing higher spin charge, whereupon the Wilson line is the higher spin entanglement entropy functional \cite{Ammon:2013hba,deBoer:2013vca}. This corresponds to choosing $R$ to be the ``Weyl representation'': in terms of the Dynkin labels, $d_i=1$ for all $i$. We write $R=\rho$, where $\rho$ is the Weyl vector of SL(N); the reason this has vanishing higher spin charge is that the weight vector $\rho$ maps to the Cartan element $L_0$, which commutes with all higher spin zero-mode generators $V_0^{s>2}$.  The boundary CFT entanglement entropy for a single interval stretching from 0 to $w$ is then, using $S_{EE}=[I_\rho]_{\rm finite}$ and equations \eqr{Iqfactorint} and \eqr{wilbint},

\e{}{S_{EE} =  k_{CS} \log \Gc_\rho + (\text{anti-holomorphic})}
with
\e{}{\Gc_\rho = \langle -\rho|e^{\Lambda}|\rho\rangle = \prod_{k=1}^{N-1} \Gc_k~.}
Recalling that we have explicitly computed $\Gc_k$ in terms of the eigenvalues of the matrix $\L$,\footnote{As we motivate below, a perturbative expansion of $S_{EE}$ is often desirable. In this situation, it is  more efficient to directly expand the matrix element $\langle -\rho|e^{\Lambda}|\rho\rangle$: not only is $|\rho\rangle$ a highest weight state, annihilated by lowering generators, but it also has vanishing higher spin charge, $V^{s>2}_0|\rho\rangle =0$. This leads to relatively simple matrix elements after perturbative expansion of $e^{\L}$. We use this technique in Section \eqr{vasiliev}.} this completes the determination of entanglement entropies in asymptotically AdS backgrounds of SL(N) higher spin gravity.

More interestingly, as we now explain, these results for general $N$ allow us to compute entanglement entropy in \hsl\ Chern-Simons theory as well, and hence in the Vasiliev theory with an infinite tower of higher spin fields. To compute \hsl\ entanglement, one would appear to need the \hsl\ Wilson line. However, we will utilize a well-known fact about \hsl, which is that it may be defined by analytic continuation of sl(N) to non-integer N: that is, 
\e{}{{\rm hs}[\l]\cong {\rm sl}(\l)}
with $\l \in \mathbb{C}$  (see, e.g., \cite{feigin, Fradkin:1990qk}). A corollary is that any rational function of sl(N) structure constants may be unambiguously continued to \hsl, simply by writing $N = \pm \l$. Crucial to this point is that sl(N) structure constants are polynomials in $N$.\footnote{Other, non-rational functions may be analytically continued on a case-by-case basis using Carlson's theorem \cite{Carlson}.} This technique has passed muster in earlier higher spin literature \cite{Gaberdiel:2011zw, Hijano:2013fja, Campoleoni:2013lma, Gaberdiel:2013jca}.

To employ this fact in the entanglement context, we develop various perturbative expansions, to be described in a moment, of the SL(N) Wilson line with $R=\rho$. At a fixed order in these expansions, the result always depends polynomially on the sl(N) structure constants, and hence on $N$. By performing the aforementioned analytic continuation, we obtain Vasiliev entanglement entropy without having to address the inherent difficulties of the yet-to-be-constructed \hsl\ Wilson line. We perform analogous computations for other representations $R$. We confirm the validity of this procedure by checking against independent computations directly in \hsl\ (and $W_{\infty}[\l]$ CFT) language.

Ideally, of course, we would like to compute entanglement entropy in Vasiliev theory non-perturbatively in the higher spin fields. However, this is a tall order given the state of affairs of 3D higher spin gravity: to do so would require resolving major conceptual issues that are present in other, more basic computations. In short, we know very little about Vasiliev theory away from perturbation theory. At present, then, the only reasonable target for Vasiliev or \hsl\ entanglement entropy computations involves a perturbative expansion in the background higher spin fields. This would be the precise entanglement parallel of the perturbative thermal partition function calculations of \cite{Kraus:2011ds,Gaberdiel:2012yb,Long:2014oxa}, and as we now describe, this is exactly the computation we have done here. It will obviously be very interesting to eventually understand non-perturbative higher spin physics in AdS$_3$. We provide a more detailed perspective on these issues in the Discussion.

In Section \ref{vasiliev}, we perform the following perturbative calculations of entanglement entropy:
\vs
$\bullet$\quad {\bf Small charges:} We expand the Wilson line perturbatively in the background higher spin charges. We explicitly demonstrate this procedure for spin-3. The result to first nontrivial order is in equation \eqr{ezc2}. One background of particular interest is the higher spin black hole. Expressing our general result \eqr{ezc2} in terms of the inverse temperature $\beta$ and chemical potential $\mu$, we arrive at the single-interval entanglement entropy in the \hsl\ black hole background of Vasiliev theory. This result passes two checks: one, it matches a CFT calculation of the same quantity \cite{Datta:2014ska,Datta:2014uxa}; and two, the large interval limit correctly reproduces the thermal entropy as computed directly in the \hsl\ Chern-Simons theory.

$\bullet$\quad {\bf Small $\mu$ in the $T=0$ higher spin black hole:} At zero temperature but nonzero chemical potential, the \hsl\ black hole solution simplifies, as does the computation of entanglement entropy in that background. The result through $O(\mu^6)$ is in equation \eqr{gzzc}. At $\l=0$, we subject our bulk calculation to a highly nontrivial test against the entanglement entropy in a CFT of $c$ complex free fermions. This theory realizes two copies of a $W_{1+\infty}$ chiral algebra, which reduces to the asymptotic $W_{\infty}[0]$ algebra of an hs[0] Chern-Simons theory in AdS$_3$ after modding out a U(1) current. This CFT calculation was initiated in \cite{Datta:2014ska}, and we extend it to higher orders here. The results agree with the Wilson line result at $\l=0$. This is a strong check of our prescription for computing Vasiliev entanglement entropy using the bulk Wilson line.

$\bullet$\quad {\bf Short interval:} We compute the leading higher spin correction to the short interval expansion of the entanglement entropy, for generic higher spin background charges. Despite this calculation being non-perturbative in the charges, the leading dependence is simple and easy to understand; the result is in equations \eqr{ezm} and \eqr{f3}.

\subsubsec{Virasoro blocks from $W_N$ blocks (Section \ref{conformalblocks})}

\begin{figure}[t!]
   \begin{center}
 \includegraphics[width = .76\textwidth]{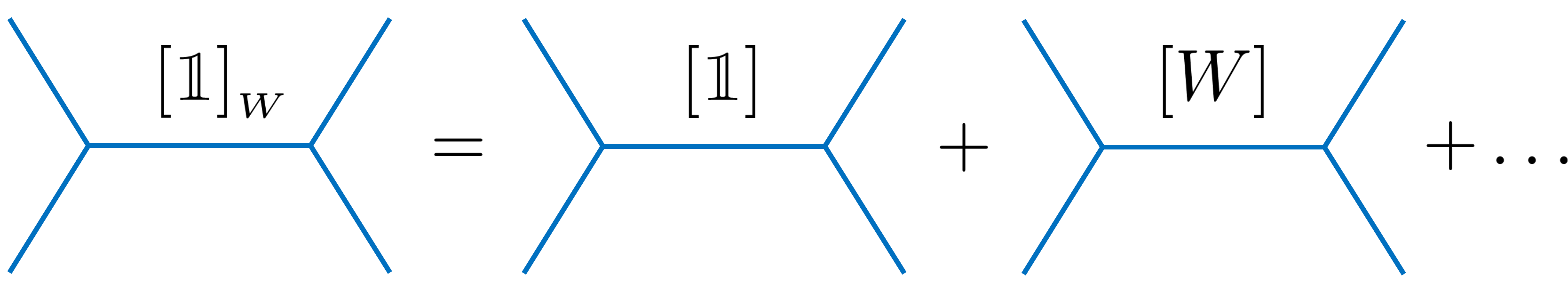}
 \caption{For external operators with higher spin charge with respect to some $W$-algebra, the  $W$-algebra vacuum block (left side) can be branched into an infinite sum of Virasoro blocks (right side). Pictured are the Virasoro blocks corresponding to vacuum and single current exchanges.}
 \label{fig}
 \end{center}
 \end{figure}

Because $W_N$ contains a Virasoro subalgebra, a $W_N$ conformal block may be branched into Virasoro conformal blocks. A corollary of this statement is that knowledge of the $W_N$ vacuum block can be used to obtain the {\it non-vacuum} Virasoro block for pairwise identical external operators. The logic is depicted in Figure \ref{fig}: allow the external operators to have spin-$s$ charges, and extract the piece of the $W_N$ vacuum block linear in these charges. 
This corresponds to the spin-$s$ current exchange, and gives the Virasoro block for exchange of an operator of dimension $s$. Although $s \in \mathbb{Z}$ so far, rationality of the Virasoro block in $s$ (at any fixed order in the cross-ratio expansion) permits the naive analytic continuation away from the integers. Using the Wilson line, we extract the semiclassical Virasoro block using the above method. The latter was recently derived in \cite{Fitzpatrick:2014vua}, and we match that result here; see equation \eqr{cfthyp}.

\subsubsec{$W_N$ minimal model data (Section \ref{wnsec})}
In Section \ref{wnsec}, we explain what our results imply about $W_N$ minimal models at large $c$. The SL(N) Vasiliev theory is holographically dual to a certain non-unitary, large $c$ limit of the $W_N$ minimal models (called the ``semiclassical'' limit in \cite{Gaberdiel:2012ku, Perlmutter:2012ds}). This permits an interpretation of our results in the context of that specific CFT.

In particular, while all of our Wilson line results are completely general and depend only on representation theory, the finite-dimensional probe representations $R$ have realizations as operators in the $W_N$ minimal models. In this way, many of our technical results may be viewed as statements about the minimal model spectrum. For instance, in Section \ref{probe charges} and Appendix \ref{probe charge}, we derive all higher spin charges of arbitrary representations $R$; see equations \eqr{gzba} and \eqr{gzna}. These then double as the charges of operators $(\L_+,0)$ in the minimal models \cite{Gaberdiel:2010pz}. In fact, according to a conjecture in \cite{Gaberdiel:2011zw}, these are also the charges of the transposed operators $(\L_+^T,0)$ in the unitary 't Hooft limit of Gaberdiel and Gopakumar \cite{Gaberdiel:2010pz}.

Our results are in line with the statement that the Wilson line action computes the vacuum block contribution to semiclassical four-point functions of SL(N) Vasiliev theory.   If one takes $c\rightarrow \infty$ while also scaling up the dimensions and charges of the external operators in proportion to $c$ (see \eqr{sclim}), then general arguments \cite{Hartman:2013mia} lead one to the conclusion that the vacuum block dominates in this limit.  In fact, a stronger statement appears to be true: explicit computations \cite{Hijano:2013fja} in the $W_N$ minimal models  show that correlators of the sort being discussed here are given by the vacuum block even if one holds fixed the dimensions/charges of one pair of external operators as $c\rightarrow \infty$.\footnote{This is known as the ``heavy-light'' limit when applied to Virasoro blocks \cite{Fitzpatrick:2015zha}.} In particular, for such operators the Wilson line result agrees exactly with the correlators computed in the non-unitary $W_N$ minimal model. The underlying reason for this seems to be that the heavy primary operators $(0,\Lambda_-)$ are dual to soliton solutions which are completely smooth in the sense of having trivial gauge holonomy.  Furthermore, only the higher spin gauge fields are excited and not any matter fields. The light probe then just sees a smooth solution built purely out of higher spin fields, which is the description one expects for the vacuum block.

\vs
\centerline{\rule{6cm}{0.4pt}}
\vs
\vs

We close the paper with a discussion of future directions and some open problems in the world of 3D Vasiliev in Section \ref{disc}. Several appendices contain technical supplements to the main text.

\section{Brief Review of 3D Higher Spin Gravity and Wilson Lines}\label{sec2}

Here we give a very brief statement of the needed facts about 3D higher spin gravity, and then about its Wilson lines.  This material is standard, and more details can be found in, e.g. \cite{Campoleoni:2010zq,Ammon:2012wc}.

A pure higher spin theory with gauge fields of spins $s=2,3,\ldots N$ is based on SL(N) $\times$ SL(N) Chern-Simons theory with action\footnote{The version that is relevant for duality to a healthy boundary CFT is instead based on the infinite-dimensional gauge group \hsl$\times$\hsl.  The SL(N) theory can be thought of as a non-unitary extrapolation, involving taking $\lambda =\pm  N$.}
\be
I=I_{CS}[A]-I_{CS}[\Ab]~,
\ee
with
\be
I_{CS}[A]={k_{cs}\over 4\pi}\int_M \Tr \left(A\wedge dA +{2\over 3}A\wedge A \wedge A\right).
\ee
We take the trace to be in the defining representation, where $(A,\Ab)$ are traceless $N\times N$ matrices.
The action is invariant under  gauge transformations
\be
\delta_\Lambda A=d\Lambda+[A,\Lambda], \quad\quad
\delta_{\overline{\Lambda}} \bar{A}=d\bar{\Lambda}+[\bar{A},\bar{\Lambda}].
\ee
The equations of motion imply that the connections are flat
\be
F=dA+A\wedge A =0, \quad\quad \bar{F}=d\bar{A}+\bar{A}\wedge \bar{A}=0.
\ee

Some basic results on SL(N) group theory are stated in   appendix \ref{group}.   Here we just note that the SL(N) generators are denoted
\be
V^s_m~,\quad s=2,3, \ldots, N~,\quad m=-(s-1), -(s-2), \ldots s-1~.
\ee
The generators $L_m \equiv V^2_m$ form a principally-embedded SL(2) subalgebra,
\be
[L_m,L_n]=(m-n)L_{m+n}~,
\ee
and $V^s_m$ transform in the spin $s-1$ representation under this SL(2),
\be
[L_m, V^s_n]=(m(s-1)-n)V^s_{m+n}~.
\ee

An asymptotically AdS$_3$ connection takes the form
\bea\label{adsbc}
A&=& \left(e^\rho L_1 + \sum_{s=2}^{N} Q_s e^{-(s-1)\rho} V^s_{-(s-1)}\right)dw + L_0 d\rho\cr
\Ab&=& \left(e^\rho L_{-1} + \sum_{s=2}^{N} \overline{Q}_s e^{-(s-1)\rho} V^s_{s-1}\right)d\wb - L_0 d\rho
\eea
where $w$ is a complex coordinate.  $(Q_s,\overline{Q}_s)$ are identified with left and right moving spin-$s$ currents, respectively.  In general, flatness permits $Q_s=Q_s(w)$ and $\overline{Q}_s=\overline{Q}_s(\wb)$, although in this paper we restrict them to be constant.
We can also include terms that correspond to adding to the dual CFT Lagrangian a source $\mu_s$ coupled to the spin-$s$ current $Q_s$:
\be\label{source}
A \rt A + \mu_s e^{(s-1)\rho} V^s_{s-1}d\wb+ \ldots
\ee
where the $\ldots$ denote additional terms required by flatness.

Given a solution described by flat connections there is an associated solution in the metric formulation. For instance, the metric is
\be
g_{\mu\nu} = {1\over \Tr (L_0L_0)}\Tr(e_\mu e_\nu)~,\quad e_\mu={1\over 2}(A_\mu-\Ab_\mu)~.
\ee
The higher spin fields are similarly obtained by taking traces of higher powers of the generalized vielbein $e_\mu$. The full action for the theory in these metric-like variables is unknown, though see \cite{Campoleoni:2012hp,Campoleoni:2014tfa}. We also note that the Chern-Simons level $k_{CS}$ is related to the Brown-Henneaux central charge as
\be
c= 12 \Tr(L_0L_0) k_{CS}= N(N^2-1)k_{CS}~,
\ee
the trace being taken in the defining representation. This should be understood as the central charge of both components of the classical $W_N \times W_N$ algebra, obtained as the asymptotic symmetry of the bulk theory with AdS$_3$ boundary conditions \ref{adsbc}.

\subsection{Wilson lines}

References  \cite{Ammon:2013hba,deBoer:2013vca}  gave two independent proposals for defining entanglement entropy in SL(N) higher spin backgrounds, each involving a kind of Wilson line for the connections $A,\Ab$ taking values in the Lie algebra of SL(N). These were later shown to be equivalent \cite{Castro:2014mza}. A more general class of Wilson lines computes vacuum conformal blocks of the $W_N$ algebra in a semiclassical limit described in \cite{deBoer:2014sna}; these are parameterized by the $N-1$ higher spin charges of a probe field moving in the background of $A,\Ab$. We now introduce these in turn. In Sections \ref{wilson line I} and \ref{closed}, which contain our main technical results, we will expand on and further clarify the relation between the two proposals  \cite{Ammon:2013hba,deBoer:2013vca}, and also explain the connection to the computation of scalar correlators carried out in \cite{Hijano:2013fja}. The latter perspective will turn out to be especially advantageous when it comes to obtaining result for general $N$.

Consider a single spacelike interval on the asymptotic boundary of some higher spin background, and let $C$ be some curve in the bulk whose endpoints connect the endpoints of the chosen interval (more precisely, $C$ is homologous to the interval).  The proposal in \cite{Ammon:2013hba} starts from the Wilson line
\be\label{wilson1}
W_{\cal R}(C)=\Tr_{\cal R}\left( {\cal P} \exp \int_C {\cal A}  \right)~.
\ee
Since the connections are flat, $W_{\cal R}(C)$ is independent of the choice of curve $C$.  ${\cal R}$ denotes a representation ${\cal R}\times {\cal R}$ of SL(N)$\times$ SL(N). The choice of ${\cal R}$ corresponds to the choice of higher spin charges for a massive field moving in the background ${\cal A}$. To compute entanglement entropy, one takes ${\cal R}$ to be the Weyl representation of SL(N), which has vanishing higher spin charges, as we justify shortly. Given this choice, the Wilson line is supposed to be related to the entanglement entropy as
\be
S_{\rm EE}=-\log (W_{\cal R}(C))~.
\ee
More precisely, we should impose a near-boundary cutoff and isolate the behavior as the cutoff is taken away.

In general, (\ref{wilson1}) is quite intractable due to the path ordering.    However, simplifications occur in the limit of very ``heavy" representations ${\cal R}$, by which we mean representations whose quadratic (and possibly higher) Casimirs becomes asymptotically large as a function of large $c$.  This is the appropriate regime for the connection with entanglement entropy as computed semiclassically in the bulk.  The bulk central charge should be large, $c \gg 1$, and the quadratic Casimir for ${\cal R}$ scales as $c_2 \sim c^2$.  The simplification at large $c$ starts with rewriting (\ref{wilson1})  using the standard quantum-mechanical  formalism for rewriting matrix elements of path-ordered exponentials in terms of path integrals.  The central charge then acts as $1/\hbar$, and so a saddle-point approximation can be used.   After the dust settles, we are left with the following prescription for computing the semiclassical Wilson line.

First, we trade the flat SL(N)$\times $ SL(N) connections $(A,\Ab)$ for SL(N) group elements as
\be\label{lr}
A(x) = L(x) d L^{-1}(x)~,\quad\Ab(x) = R^{-1}(x) d R(x)~.
\ee
This can always be done locally, and nontrivial gauge holonomies can be incorporated by allowing $(L,R)$ to be multivalued.      We then define the object $M$ as
\bea\label{lrm}
M& \equiv &   {\cal P} \exp \left(-\int_{C_r}A \right)  {\cal P} \exp \left(-\int_{C}\Ab \right)     \cr
&=& L(x_i)L^{-1}(x_f)R^{-1}(x_f)R(x_i)~,
\eea
Here $C$ is the contour starting at $x_i$ and ending at $x_f$, while $C_r$ denotes the reversed contour, starting at $x_f$ and ending at $x_i$.   $M$ is the ``composite Wilson line" employed in \cite{deBoer:2013vca}.  Now consider $M$ in the defining $N\times N$ matrix representation, and let $\lambda_M$ denote the diagonal matrix whose entries are the eigenvalues of $M$, 
\e{}{\lambda_M = \text{diag}(\text{eig}(M))~.}
The eigenvalues should be put in the ``primary ordering" \cite{Ammon:2013hba,Castro:2014mza}. As made explicit below, $\lambda_M$ contains all the needed data characterizing the background and the contour $C$.

Next, we need to characterize the invariant data of the Wilson line probe, namely its higher spin charges. We do this by specifying an element of the SL(N) Cartan subalgebra, denoted as $P_0$:\footnote{We derive this -- specifically, the normalization of $q_s$ -- in Appendix \ref{P_0}.}
\e{probe1}{P_0 = \sum_{s=2}^N {q_s\over \Tr_{\square}(V^s_0V^s_0)}V^s_0}
$q_s$ is the spin-$s$ charge carried by the probe, with a factor of $k_{CS}$ pulled out, and the trace is taken in the defining representation, $\square$. Equivalently, we can specify a weight vector, or ``charge vector," $q$ via
\e{probe2}{P_0 = q\cdot H~.}
Again working in the defining representation so that $P_0$ is a diagonal $N\times N$ matrix, the semiclassical Wilson line is given as
\be\label{wilson2}
W_q(C) = e^{-I_q(C)}~,\quad I_q(C) = k_{CS}\Tr[\log(\lambda_M)P_0]~.
\ee
This completes the prescription.

As was just noted, entanglement entropy is given by a particular choice for $P_0$, or equivalently $q$.  The precise choice for $P_0$ can be motivated by consideration of the replica trick, where we can think of the Wilson line as a particle that creates a conical defect in the metric.  Importantly, to compute entanglement entropy there should only be a conical defect in the metric, and not in the higher spin fields.   In terms of our explicit SL(N) Cartan generators, this means that $P_0 \propto L_0$.  The coefficient of proportionality can also be motivated by the replica trick, or by matching to the known result for empty AdS$_3$.  Either way, one has
\be\label{Pdef}
P_0 =  L_0 \quad\quad {\rm (entanglement ~entropy)~.}
\ee
The corresponding charge vector is $q=\rho$, where $\rho$ is the Weyl vector. This describes the representation ${\cal R}$ whose Young diagram is ``triangular,'' with $N-i$ boxes in the $i$'th row.

\subsubsec{$W_N$ vacuum blocks and four-point functions}
In an AdS/CFT context, the SL(N) Wilson line is computing a boundary two-point function in an excited state. Consider a pair of CFT operators $\Oc_H$ and $\Oc_L$,  primary with respect to currents $J^{(s)}$ that generate the $W_N$ algebra:
\e{hlcharges}{J^{(s)}|\Oc_H\rangle = k_{CS}Q_s|\Oc_H\rangle~, \quad J^{(s)}|\Oc_L\rangle = k_{CS} q_s|\Oc_L\rangle}
We have intentionally used the same notation for the charges $Q_s$ and $q_s$ as in \eqr{adsbc} and \eqr{probe1}, respectively. We want to consider their four-point function on the cylinder,
\e{hllh}{\langle \Oc_H(-\infty)\Oc_L(0,0)\Oc_L(w,\wb)\Oc_H(\infty)\rangle = \langle \Oc_H|\Oc_L(0,0)\Oc_L(w,\wb)|\Oc_H\rangle~, }
in the following large $c$ regime:
\e{sclim}{c\rar\infty~, \quad Q_s, q_s~\text{fixed}~,\quad q_s\ll 1~.}
This is the limit to which we refer throughout most of this paper; in line with previously established terminology, we will call this the ``semiclassical'' limit.\footnote{Taking $q_s$ to not be small is also a kind of semiclassical limit that we will not discuss here; our use of ``semiclassical'' is in the restricted sense \eqr{sclim}.} The correspondence of the four-point function to the Wilson line follows from application of the AdS/CFT dictionary: the ``heavy'' operators $\Oc_H$ set up the higher spin background in the bulk through $A,\Ab$, while the ``light'' operators $\Oc_L$ map to a massive, charged particle carrying $\Oc_L$'s quantum numbers that traces out a bulk worldline. In this way, the Wilson line is naturally interpreted as the light two-point function in this heavy background.

Moreover, the near-boundary limit of SL(N) Wilson lines with general charge vectors computes the semiclassical vacuum conformal blocks of the $W_N$ algebra \cite{deBoer:2014sna}. The block is given by the solution of a differential equation subject to a monodromy prescription for an auxiliary SL(N) Chern-Simons connection. The Wilson line for this connection, now regarded as that of a bulk higher spin theory, computes the same quantity. This was explained in \cite{deBoer:2014sna} and shown by brute force computation for $N=3$.

Following the same line of argument as in the Virasoro case \cite{Hartman:2013mia}, we expect that the semiclassical four-point functions described above are exponentially dominated by the vacuum $W_N$ conformal block in the semiclassical limit, which we denote ${\cal F}_{\rm vac}^{\,W_N}$. So the descriptions of the Wilson line as both $W_N$ vacuum block and correlator hold together, and may be summarized as
\e{wblock}{W_{\cal R}(C) = |{\cal F}_{\rm vac}^{\,W_N}|^2~.}
As with all statements about CFT quantities throughout this paper, the Wilson line is to be evaluated in the SL(N) theory in the near-boundary limit; the matching of probe and light operator quantum numbers in \eqr{wblock} is implied.

Henceforth we will consider a general charge vector, although we occasionally comment on the special case relevant for entanglement entropy.

\section{SL(N) Wilson Line I. A Simpler Formulation}\label{wilson line I}
In this section, the first of two containing our main formal results, we give new and simple expressions for the near-boundary limit of a general Wilson line in asymptotically AdS$_3$ backgrounds of SL(N) higher spin gravity. 

The expression (\ref{wilson2}) requires us to compute the eigenvalues of $M$.  In this paper we will restrict to connections of the form\footnote{The radial coordinate $\rho$ is not to be confused with the Weyl vector; after this Section, the radial coordinate makes no further appearance.}
\bea
A &= & b^{-1} a b +b^{-1} db~,\quad  a=a_w dw + a_{\wb}d\wb\cr
\Ab &= & b \ab b^{-1} +b db^{-1}~,\quad  \ab=\ab_w dw + \ab_{\wb}d\wb\cr
b&=& e^{\rho L_0}~,
\eea
with $a_{w,\wb}$ and $\ab_{w,\wb}$ constant.  Flatness requires $[a_w,a_{\wb}]=[\ab_w,\ab_{\wb}]=0$.  In this case, the gauge functions in \eqr{lr} read
\be
 L = e^{-\rho L_0}e^{-a_w w -a_{\wb}\wb}~,\quad R = e^{\ab_w w + \ab_{\wb}\wb}e^{-\rho L_0}~.
\ee
For the endpoints of the contour we take $x_{i,f}=(\rho, w, \wb)_{i,f}$ with
\be
x_i = (-\log \eps, 0,0)~,\quad  x_f=(-\log \eps, w, \wb)~,
\ee
where we used translation invariance to set $w_i=0$. This gives, using \eqr{lrm},
\be\label{Mform}
M =e^{\log \eps L_0}e^{a_w w+a_{\wb}\wb}e^{-2\log \eps L_0}e^{-\ab_w w-\ab_{\wb}\wb}e^{\log \eps L_0}~.
\ee

It will turn out to be convenient to write  $\lambda_M$ in terms of parameters $\gamma_j$ as
\be\label{lampar}
\lambda_M= \exp \left[ \sum_j \log(\gamma_j)\alpha^{(j)}\cdot H \right]~.
\ee
Here $\alpha^{(j)}$ are the simple roots of the SL(N) algebra.  Recall that the simple roots obey $\alpha^{(i)}\cdot \omega^{(j)} = \delta_{ij}$, where $\omega^{(j)}$ are the fundamental weights.

We consider $P_0$ of the form
\be\label{P0d}
P_0 = d \cdot H~,\quad  d = \sum_i d_i \omega^{(i)}~.
\ee
In general, $d_i\in \mathbb{R}$. When $d_i \in \mathbb{Z}^+$, as is the case for finite-dimensional representations, the $d_i$ are the Dynkin labels associated to the Cartan element $P_0$. Otherwise, they are associated to infinite-dimensional representations.\footnote{A useful guide to keep in mind is the case of SU(2). There is one Dynkin label, $d_1$, which equals twice the spin $j$ of a given representation: $d_1=2j$. When $j \in {1\over 2} \mathbb{N}$, the representation is $(2j+1)$-dimensional, and $d_1$ equals the number of boxes of a Young diagram. For $j \notin {1\over 2} \mathbb{N}$, we have $d_1\notin \mathbb{Z}$ and the representation is infinite-dimensional.} We then find
\be\label{Iq}
I_q = k_{CS}\log\left( \gamma_1^{d_1} \gamma_2^{d_2} \ldots \gamma_{N-1}^{d_{N-1}}\right)~.
\ee
This is a key formula. Of course, in this expression we should extract the behavior as $\eps \rightarrow 0$ to obtain the physically relevant Wilson line result.

The relation between the $d_k$ and the higher spin charges $q_s$ -- that is, the bases \eqr{probe1} and \eqr{P0d} -- is
\e{}{d_k = (P_0)_{kk} - (P_0)_{k+1,k+1}~.}
This allows one to compute the $d_k$, given some set of higher spin charges $q_s$. In Section \ref{probe charges} we will use another method to invert this relation, obtaining the reverse map $d_k \mapsto q_s$.

Let's briefly restrict to the special case relevant for entanglement entropy.  Recall that the Weyl vector $\rho$ has Dynkin labels $d_1= d_2 = \ldots = d_{N-1}=1$.   This gives
\be
S_{\rm EE} =  k_{CS}\Tr[\log(\lambda_M) L_0] = k_{CS}\log\left( \gamma_1 \gamma_2 \ldots \gamma_{N-1}\right)~,
\ee
which is correct \cite{Castro:2014mza}.

We now explore the further simplifications for asymptotically AdS$_3$ connections. For orientation, it's useful to first consider the case of empty AdS$_3$, which will also serve as a check of the normalization in (\ref{Pdef}).
Empty AdS$_3$ with planar boundary is represented by the connection
\be
a= L_1 dw~,\quad \ab=L_{-1}d\wb.
\ee
This gives
\be
M=  e^{{1\over \eps}L_1 w}e^{-{1\over \eps}L_{-1}\wb}~.
\ee
This is a product of SL(2) group elements, and hence is conjugate to some group element of the form $e^{\alpha L_0}$.    The constant $\alpha$ is easily computed by computing and comparing eigenvalues in the $2\times 2$ representation of SL(2), and this yields
\be\label{lamM}
\lambda_M= e^{2 \log \left(w\wb\over \eps^2\right)L_0}~.
\ee
This gives the entanglement entropy
\be
S_{\rm EE} = 2k_{CS}\Tr(L_0 L_0) \log \left(w\wb\over \eps^2\right) = {c\over 6} \log\left(w\wb\over \eps^2\right)~,
\ee
which is the standard result.

From (\ref{lamM}) we may read off the various powers of $\eps$ carried by the eigenvalues
\be\label{eigs}
\lambda_M = {\rm diag}\left(c_1 \eps^{-2(N-1)}, c_2 \eps^{-2(N-3)}~,\ldots , c_N \eps^{2(N-1)} \right)~,
\ee
with coefficients $c_i = (w\wb)^{N+1-2i}$. In terms of the $\gamma_i$,
\e{cm}{(\lambda_M)_{ii} = {\gamma_i\over \gamma_{i-1}}~, \quad \text{where} \quad \gamma_0 = \gamma_N \equiv 1~.}
A key point is as follows. We will  be considering connections more general than empty AdS$_3$; what changes in those cases are the coefficients $c_i$, while the relation \eqr{cm} and the leading powers of $\eps$ in each eigenvalue remain as in (\ref{eigs}).    This is true for any asymptotically AdS$_3$ connection.  Actually, we will also be considering some non-asymptotically AdS$_3$ connections, which could alter this conclusion.  However, in those cases we will always do perturbation theory in the parameters that change the asymptotics, and this effectively brings us back to the asymptotically AdS$_3$ case.

In general, the Wilson line action will have the structure
\be
I = {\rm (universal~constant)}\times \log \eps + {\rm (finite)} + {\rm (vanishing ~as~}\eps \rt 0)~.
\ee
The universal constant takes the same value for all backgrounds and for all intervals, whereas the finite piece depends on the background and on the interval size.   We are only interested in this background/interval-dependent contribution to the finite part, since we can always modify the constant part by rescaling the cutoff $\eps$.   With this in mind, we will at times freely drop such constant finite parts in $I$.

Now we give another useful expression for the Wilson line action. As discussed above, for pure AdS $\lambda_M$ is given by (\ref{lamM}).   It's clear that $\Tr (\lambda_M)$ is dominated, as $\eps \rt 0$, by the state with highest $L_0$ eigenvalue; we henceforth refer to this as the highest weight state $|{\rm hw}\rangle$.  This statement is also true for a more general asymptotically AdS connection, or in perturbation theory around such a connection.   Non-highest weight states will give contributions suppressed by positive powers of $\eps$ compared to the highest weight state.

Let us consider $M$ in a general highest weight representation $R$ with highest weight state $|{\rm hw}_R\rangle$; we write $M$ in this case, and $\lambda_{M}$ for its diagonal form.    For pure AdS, \eqr{lamM} again holds, where $L_0$ now denotes the Cartan generator $\rho\cdot H$ in the representation $R$.  By the same logic as above, for a general asymptotically AdS connection we have
\be
\Tr_R  [M] =\Tr_R[\lambda_{M}] =  \langle {\rm hw}_R|\lambda_{M}|{\rm hw}_R\rangle\quad {\rm as } \quad \eps\rt 0~.
\ee
To obtain an explicit expression, expand the highest weight in terms of the fundamental weights, hw$_R = \sum_i d_i \omega^{(i)}$, and use the parametrization (\ref{lampar})  (in representation $R$). This yields
\be
\langle {\rm hw}_R|\lambda_{M}|{\rm hw}_R\rangle = \gamma_1^{d_1}\gamma_2^{d_2} \ldots \gamma_{N-1}^{d_{N-1}}~.
\ee
Comparing to (\ref{Iq}) we obtain
\be\label{Iqres}
I_q = k_{CS} \log \Tr_R  [M] \quad {\rm as } \quad  \eps\rt 0~.
\ee
Note that this applies for all highest weight representations $R$, whether finite- or infinite-dimensional.

\subsection{Factorization}

The above expressions are still rather inconvenient for explicit computation, for two reasons.   First of all, one has to compute the Wilson line action and then extract the $\eps \rt 0$ behavior.  It would be more convenient to have a general expression that gives the $\eps \rt 0$ contribution directly.     Second, as we will see momentarily,  in this limit the Wilson line factorizes into parts determined by $A$ and $\Ab$ respectively.  It is much simpler to compute each of  these factors individually, but it is not clear how to do this using the expressions above.

We now give a result for the Wilson line that rectifies these two shortcomings for arbitrary probe charges.

Let's first consider finite-dimensional representations $R$. These contain both highest and lowest weight states. Let $h_R$ be the highest eigenvalue of $L_0$ in representation $R$, and $-h_R$ the corresponding lowest weight. That is, take the highest and lowest weight states to obey
\e{}{L_0 \ket{\pm \hw_R} = \pm h_R \ket{\pm \hw_R}~.}
The highest and lowest weight conditions are
\e{hwlw}{V^s_{n<0}\ket{\hw_R} =0~,\quad
V^s_{n>0}\ket{-\hw_R}=0~.}
Projectors onto these highest/lowest weight states are
\be
P_\pm   =  \ket{\pm \hw_R} \bra{\pm \hw_R} = \lim_{\eps \rt 0}  \eps^{2h_R} e^{\mp 2 \log \eps L_0}~,
\ee
and so, from \eqr{Iqres},
\bea
I_q &=&k_{CS} \log \left\{\eps^{-4h_R} \Tr_R \left[ P_- e^{a_w w+a_{\wb}\wb}P_+e^{-\ab_w w-\ab_{\wb}\wb}\right]\right\}\cr
& = &k_{CS} \log \left\{\eps^{-4h_R} \langle -{\rm hw}_R| e^{a_w w+a_{\wb}\wb} | {\rm hw}_R\rangle \langle {\rm hw}_R| e^{-\ab_w w-\ab_{\wb}\wb} \ket{-{\rm hw}_R} \right\}  ~,
\eea
as $\eps \rt 0$.    The interesting finite part thus reduces to a sum of two terms, each of which depends only on one of the connections,
\be\label{Iqfactor}
[I_q]_{\rm finite} =k_{CS} \log  \langle -{\rm hw}_R| e^{a_w w+a_{\wb}\wb} | {\rm hw}_R\rangle+k_{CS} \log \langle {\rm hw}_R| e^{-\ab_w w-\ab_{\wb}\wb} \ket{-{\rm hw}_R}~.
\ee
As promised, we now have a convenient expression with the $\eps$ dependence stripped off, and separated into contributions from the two connections.

The extension to include infinite-dimensional representations is simple. Starting from \eqr{Iq}, a convenient way to write the general Wilson line $I_q$ is by using properties of the antisymmetric tensor representations $R = \asym_k$. These have $d_i=\delta_{i,k}$, and highest and lowest weight states which we denote $\ket{\hw}_k$ and $\ket{-\hw}_k$, respectively. Their Wilson line action, which we denote $I_k$, is
\e{}{I_k = k_{CS} \log \gamma_k~. }
The general Wilson line action is then
\e{Iqk}{I_q = \sum_{k=1}^{N-1} d_k I_k }
where $d_k$ are arbitrary real numbers parameterizing an arbitrary representation $R$. Since the $\asym_k$ representations are themselves finite-dimensional, the above result \eqr{Iqfactor} applies to each $I_k$ individually. Plugging these into \eqr{Iqk} then gives the general result in factorized form:
\be\label{Iqfactor2}
[I_q]_{\rm finite} =k_{CS}\log \Gc_R + k_{CS}\log \overline{\Gc_R}
\ee
where
\e{Gform}{\Gc_R \equiv \prod_{k=1}^{N-1} \bra{-\hw} e^{a_w w+a_{\wb}\wb} | \hw\rangle_k^{d_k} \quad \text{and}\quad \overline{\Gc_R} \equiv \prod_{k=1}^{N-1} \bra{\hw} e^{-\ab_w w-\ab_{\wb}\wb} \ket{-\hw}_k^{d_k}~.}
This is one of our main results. In short, to compute the Wilson line in the desired asymptotic limit $\eps\rar0$, one need only compute chiral matrix elements in the $\asym_k$ representations, and multiply them as above. In the next section, we will compute these explicitly in terms of the eigenvalues of the connections $(a,\ab)$.

\section{SL(N) Wilson Line II. Explicit Expression}\label{closed}

To summarize where we are so far, after stripping off a universal divergent part, the  Wilson line action $I_q = -\log W_q$ is given by
\be\label{wila}
[I_q]_{\rm finite} =k_{CS} \log \Gc_R + k_{CS}\log \overline{\Gc_R}~,
\ee
with
\e{wilb}{\Gc_R = \prod_{k=1}^{N-1}\Gc_k^{\,d_k}~,\quad \overline{\Gc_R}=\prod_{k=1}^{N-1}\overline{\Gc_k}^{\,d_k}}
where
\e{wilc}{\Gc_k \equiv \bra{-\hw}e^{\Lambda}|\hw\rangle_k~,\quad \overline{\Gc_k}\equiv \langle \hw|e^{-\overline{\Lambda}}\ket{-\hw}_k}
and $(\Lambda, \overline{\Lambda})$ are defined as in \cite{Hijano:2013fja}
\be\label{lamb}
\Lambda \equiv a_w w +a_{\wb}\wb~,~~\overline\Lambda \equiv \ab_w w + \ab_{\wb}\wb~.
\ee
The probe charges are encoded in the representation $R$, which has highest weight vector hw$_R=d$, where $d_i$ denote the (generalized) Dynkin labels, and charge vector $q= d$. The background charges are encoded in $(\L,\overline{\L})$, and hence in $(\Gc_k, \overline{\Gc_k})$.

In this section, we will consider a  general asymptotically AdS$_3$ connection.   Such a connection can always be put into ``highest weight gauge", for which
\bea\label{hwhw}
\Lambda &=& a_w w = \left(L_1+ \sum_{s=2}^{N} Q_s V^s_{-(s-1)}
\right)w\cr
\overline{\Lambda} &=& \ab_w \wb = \left(L_{-1}+ \sum_{s=2}^{N} \overline{Q}_s V^s_{(s-1)}
\right)\wb~.
\eea
Given such a connection, our task is to compute $\Gc_R$ and $\overline{\Gc_R}$ for a general SL(N) representation $R$.  In the following we focus on $\Gc_R$; the result for $\overline{\Gc_R}$ is obtained from this by making the replacements
\be
Q_s \rt (-1)^s \overline{Q}_s~,\quad  w \rt \wb~.
\ee
We also drop the subscript from $[I_q]_{\rm finite}$, and from now on focus on this piece exclusively.

\subsection{Computation of $\Gc_R$}\label{compG}

The approach  followed here is an extension of that in \cite{Hijano:2013fja}, which in turn uses ideas in \cite{Castro:2011iw}. Our work is reduced to computing $\Gc_k(w)$, where we have indicated the holomorphic dependence on $w$ due to the form of the connections \eqr{hwhw}.  This was done in \cite{Hijano:2013fja} for $k=1,2$; here we generalize to arbitrary $k$.

The result will be expressed in terms of the eigenvalues of $\Lambda$ in the defining representation,
\be
{\rm eig}(-\Lambda) =  (\lambda_1, \lambda_2, \ldots \lambda_N)~.
\ee
In the defining representation we write the states as $|i\rangle$, with  highest weight state  $|1\rangle$ and  lowest weight state  $|N\rangle$.
For the k-fold antisymmetric product $\asym_k$, the highest and lowest weight states are
\bea|{\rm hw}\rangle_k &=& {1\over \sqrt{k!}} \eps_{i_1 \ldots i_k} |i_1\rangle \ldots |i_k \rangle~, \cr  \ket{-{\rm hw}}_k &=& {1\over \sqrt{k!}} \eps_{i_1 \ldots i_k} |N+1-i_1\rangle \ldots |N+1-i_k \rangle~.
\eea
This implies
\be\label{detform}
 \langle-{\rm hw}| e^{\Lambda} |{\rm hw}\rangle_k = \det X
\ee
where $X$ is a $k\times k$ matrix with entries
\be
X_{mn}= \langle N+1-m | e^{\Lambda} |n\rangle~,\quad    m, n=1, 2, \ldots k~.
\ee

In appendix \ref{SLNappendix} we show how to compute $\det X$ in terms of the eigenvalues $\lambda_i$.  This leads to the result
\bea\label{Gkresult}
&&\Gc_k(w)=\langle-{\rm hw}| e^{\Lambda} |{\rm hw}\rangle_k  =
\cr &&(-1)^{k(k-1)/2}\left[\prod_{j=1}^k {\Gamma(N+1-j)\over \Gamma(j+1)}\right]\sum_{j_1, j_2, \ldots j_k=1}^N V[\lambda_{j_1}, \lambda_{j_2}, \ldots \lambda_{j_k}]^2    \left[\prod_{a_1 \neq j_1} {1 \over \lambda_{j_1} - \lambda_{a_1} }\right] \left[\prod_{a_2 \neq j_2} {1 \over \lambda_{j_2} - \lambda_{a_2} }\right]\cr
&& \quad\quad\quad\quad\quad\quad\quad\quad\quad\quad\quad
\quad\quad\quad\quad\quad\quad\quad\quad \ldots  \left[\prod_{a_k \neq j_k} {1 \over \lambda_{j_k} - \lambda_{a_k} }\right]e^{-\lambda_{j_1}}\ldots e^{-\lambda_{j_k}}\cr
&&  ~,
\eea
with
\begin{align}\label{V define}
V[\lambda_{j_1},\ldots,\lambda_{j_k}]  =
\begin{vmatrix}
\lambda^{k-1}_{j_1} & \ldots & \lambda^{k-1}_{j_k}\\
\vdots & \ddots & \vdots\\
\lambda_{j_1} & \ldots & \lambda_{j_k}\\
1 & \ldots & 1
\end{vmatrix}~.
\end{align}
This is another of our main results. Using equations \eqr{wila}--\eqr{wilc}, we now have an expression for the general Wilson line evaluated in an asymptotically AdS$_3$ connection.

\subsection{A check: thermal entropy from the large interval limit}

Upon taking the size of the interval to infinity, the entanglement entropy should reduce to the thermal entropy.  We now verify that the our general result exhibits this property.

In the  large interval limit, the eigenvalues $\lambda_i$ all grow proportionally to the interval size $L$.  The right hand side of \ref{Gkresult} is then dominated by the term in the sum involving the $k$ smallest eigenvalues; other terms are exponentially suppressed.  Let us order the eigenvalues as $\lambda_1 < \lambda_2 < \ldots < \lambda_N$,   so we can write
\bea
\log \Gc_k(w) =-\sum_{i=1}^k \lambda_i + {\rm (non-extensive)}
\eea
as $L\rightarrow \infty$.    From \ref{wila}--\ref{wilb} we have for the extensive
term, in the Weyl representation $(d_i=1)$,
\bea
I_\rho &=& -k_{CS}\sum_{k=1}^{N-1} \log \Gc_k(w)=-k_{CS}\sum_{k=1}^{N-1}\sum_{i=1}^k \lambda_i \cr
 &=&-k_{CS} \left[ (N-1)\lambda_1 + (N-2)\lambda_2+ \ldots + \lambda_{N-1}\right]~,
\eea
where we suppress the corresponding $\overline{\Gc}$ term.  Recalling that the eigenvalues sum to zero, this can equally well be written
\bea
I_\rho & = & -k_{CS} \left[ {N-1\over 2} \lambda_1 + {N-3\over 2} \lambda_2+ \ldots - {N-1\over 2} \lambda_{N}\right] \cr
& = & -k_{CS}\Tr [L_0 \lambda]~,
\eea
where we recall the result \ref{L0form} for $L_0$ in the defining representation.   The entanglement entropy then becomes
\bea
S_{EE} = - I_\rho = k_{CS}\Tr [L_0 \lambda]~,
\eea
which reproduces the form of the thermal entropy first written in \cite{deBoer:2013gz}.

We conclude this section with a comment about gauge invariance.  Applied to a closed loop in a solution with spatially periodic boundary conditions, the eigenvalues $\lambda_i$ are fully gauge invariant, as they are the eigenvalues of the holonomy around a closed loop. Hence the entropy assigned to a black hole using this expression is independent of gauge choice.   On the other hand, for entanglement entropy we use an open Wilson line, which is not gauge invariant under gauge transformations that are nonvanishing at the endpoints of the interval.   This leads to subtleties  related to the choice of ``holomorphic" versus ``canonical" prescriptions which are not entirely understood, as we will note in the following when comparing bulk and CFT results;  see also the comments in appendix \ref{prescrips}.

\section{SL(N) Wilson Lines from SL(N) Vasiliev Theory}\label{vasilievwilsonlines}
We now pause to note that the result (\ref{Iqfactor})  connects onto the computation of scalar correlators in Vasiliev theory, as performed in \cite{Hijano:2013fja}. This yields a better understanding of why the Wilson line yields sensible answers, as we will show how it arises from  the perturbative expansion of a full higher spin theory coupling gauge fields to matter --- specifically, the Vasiliev theory with SL(N) gauge fields.   We also  argue that the vacuum conformal block dominates heavy-light four-point functions in the class of coset CFTs with Vasiliev duals. We now explain this, starting with a few words about \cite{Hijano:2013fja}.

The work \cite{Hijano:2013fja} was concerned with the computation of four-point functions in Vasiliev's higher spin gravity. This theory contains a free parameter $\l$. At $\lambda=\pm N$ where $N\in \mathbb{Z}$, the pure higher spin sector reduces to two copies of SL(N) Chern-Simons theory; the full Vasiliev theory may be regarded as implementing a gauge-invariant coupling of the higher spin gauge fields to scalar matter. This is the only known example of an SL(N) theory of higher spins coupled gauge-invariantly to matter. Via AdS/CFT, this theory corresponds to a particular non-unitary limit (known as the ``semiclassical limit'') of the coset CFTs that appear in the duality conjecture of Gaberdiel and Gopakumar \cite{Gaberdiel:2010pz, Gaberdiel:2012ku}; we review some details of this duality in Section \ref{wnsec}. Non-unitarity notwithstanding, this is an instance of higher spin holography whose bulk-boundary map is well-understood \cite{Perlmutter:2012ds}.

The particular four-point functions considered in \cite{Hijano:2013fja} were those with two identical heavy defect operators (dimensions scaling like large $c$), and two identical light scalar operators (dimensions of order $c^0$). This is the same sort of correlator discussed in equation \eqr{hllh}, except that the scalar operators here have dimensions that do not scale with large $c$ -- they are ``truly'' light. We call this regime the ``heavy-light'' regime, following \cite{Fitzpatrick:2015zha}. In the bulk, this vacuum four-point function becomes equivalent to the scalar two-point function computed in a higher spin background.\footnote{That this is the full contribution to the four-point function may be justified by noting that  the higher spin background representing a minimal model primary $(0,\Lambda_-)$ is a completely smooth solution in the sense of having trivial gauge holonomy.  There are thus no additional diagrams corresponding to the exchange of matter fields between the probe and the background; this is to be contrasted to the case where the background is replaced by a particle or singular defect solution, in which case there is a preferred location for emission/absorption of matter quanta. } These correlators were computed from the Vasiliev field equations, and precise agreement was found with the dual CFT correlators computed using the Coulomb gas approach to the coset models.

Scalar fields in Vasiliev theory are associated with some representation $R$ of SL(N).   The ``basic" scalar field in the Vasiliev theory corresponds to the defining representation, but in \cite{Hijano:2013fja} more general representations were considered and given a physical interpretation as bulk duals to multi-trace operators. For all $R$, the Vasiliev correlation function was obtained in the standard way, as the boundary limit of the scalar bulk-to-boundary propagator evaluated in the higher spin background. They were found to holomorphically factorize. We refer the reader to \cite{Hijano:2013fja} for a complete discussion. When all is said and done, the half of the scalar correlator depending on a constant unbarred connection is given simply as\footnote{In \cite{Hijano:2013fja} the sign of $\Lambda$ is reversed compared to what appears in (\ref{Gr}).  However, this is just a convention issue, as the sign can be flipped by interchanging $A \leftrightarrow \Ab$ in the equation for $C$ upon which the approach of \cite{Hijano:2013fja} is based.}
\be\label{Gr}
\Gc_R=\langle-{\rm hw}_R|e^{\Lambda}|{\rm hw}_R\rangle~.
\ee
Likewise, the anti-holomorphic half was found to be
\e{Grb}{\overline{\Gc_R}=\langle{\rm hw}_R|e^{-\overline{\Lambda}}\ket{-{\rm hw}_R}~.}
where $\Lambda$ and $\overline\Lambda$ are defined in \eqref{lamb}. The connection to the Wilson line action in (\ref{Iqfactor}) is immediate.  

Let us now try to explain the underlying reasons for this connection, which at first seems a bit mysterious.   In the large $c$ limit the Wilson line is dual to an operator whose scaling dimension and charges grow with $c$, but are still assumed to be numerically small compared to $c$ such that we can work to first order in the ratio.  On the other hand, a perturbative scalar field is dual to an operator whose scaling dimension and charges are fixed in the large $c$ limit.  These are distinct limits, but we have seen explicitly that the results from the Wilson line and the scalar correlator are equivalent.

We first note that it has been independently shown in \cite{deBoer:2014sna} that, as recalled in equation \eqr{wblock}, the Wilson line is also equal to the vacuum block of the $W_N$ algebra in the semiclassical limit \eqr{sclim}. One implication of this is that the holographic two-point function of the Vasiliev theory is itself apparently equal to the semiclassical vacuum block at leading order in large $c$. That is, the semiclassical four-point functions in the dual CFT are dominated by the exchange of the operators in the $W_N$ identity module. This explains why the bulk result in \cite{Hijano:2013fja} holomorphically factorizes.

From this perspective, our calculations have a lot in common with recent work relating correlators in ordinary gravity to properties of semiclassical Virasoro blocks \cite{Fitzpatrick:2014vua,Fitzpatrick:2015zha,Hijano:2015rla,Hijano:2015qja}. As recalled in the introduction, one can read off the semiclassical Virasoro vacuum block from the bulk scalar two-point function in a locally AdS$_3$ background. In direct analogy, one may view our results herein as an extraction of the semiclassical $W_N$ vacuum block, and hence the Wilson line, from the correlators computed in the higher spin backgrounds in \cite{Hijano:2013fja}.\footnote{This is true for the finite-dimensional representations $R$. From this perspective, the result for arbitrary representations may be viewed as an analytic continuation of the Dynkin labels $d_i$ to the reals.}

With this understanding in hand, the agreement between the two distinct limits yielding either the Wilson line or the perturbative correlator is less surprising, as it just a $W_N$ extension of something already understood in the Virasoro case.    In the Virasoro case, the explicit result for the vacuum block shows that in the case of two identical light operators the two limits are equivalent.  First, one argues \cite{Fitzpatrick:2015zha} for an equivalence between holding $h/c$ fixed and working to first order in $h/c\ll 1$, and holding $h$ fixed and working to leading order in $h \gg 1$.  Second, the explicit form of the Virasoro blocks shows that when the external operator dimensions are equal in pairs, the dependence on $h$ is so simple that working to leading order in $h \gg 1$ is in fact exact for all fixed $h$.   Putting these facts together establishes the equivalence between the $h \sim O(1)$ perturbative scalar case, and the $h/c \ll 1$ Wilson line case.   While the analogous statements have not been proven for the $W_N$ block, our results suggest that they continue to hold.

There is substantial evidence that vacuum dominance of semiclassical four-point functions is a characteristic, or perhaps even a diagnostic, of sparse, large $c$ CFTs \cite{Hartman:2013mia}.   The usual argument for this requires that the dimensions/charges of all external operators become large with $c$.  Here we have found examples for which this is true even when we hold fixed the dimensions/charges of one pair of external operators as we take $c\rightarrow \infty$.   We have shown explicitly how this works for the duality between Vasiliev theory at $\lambda=\pm N$ and the semiclassical limit of the $W_N$ minimal models.

 In Section \ref{wnsec}, we will elaborate on the role of the above in the duality between Vasiliev theory and the $W_N$ minimal models \cite{Gaberdiel:2010pz}.

\section{Entanglement in \hsl\ Vasiliev Theory, and Other Applications}\label{vasiliev}
With our expressions for the Wilson line in hand, we can now compute in various cases of interest.
\subsection{Small charge expansion of entanglement entropy}\label{smallcharge}

In this section, we investigate the case where a pure SL(2) connection is perturbed by a small spin-$s$ charge. This corresponds to a higher spin perturbation of a pure-metric conical defect or Euclidean BTZ black hole metric. Our efforts will lead to a determination of the single interval entanglement entropy in the higher spin black hole background of the \hsl\ theory of higher spin gravity, and hence in Vasiliev theory.

Given the result \eqref{Gkresult} for a Wilson line with general probe charges in an asymptotically AdS background, the most obvious way to develop a perturbative expansion is to expand the eigenvalues $\l_i$, and plug into \eqref{Gkresult}. A slicker method for finite-dimensional representations $R$, which we use here, is to work directly with the form \eqref{Gform}: in particular, the highest  and lowest weight conditions \eqr{hwlw} kill many terms appearing in a given matrix element. This is especially true for the Weyl representation required for calculating entanglement entropy, which obeys the further null conditions 
\begin{align}
\label{ea}
V^2_0\ket{\rho} = h_\rho\ket{\rho}~,~~~ V^{s>2}_0\ket{\rho} = 0
\end{align}
with
\e{}{h_\rho = {N(N^2-1)\over 12} = {c\over 12k_{CS}}}
hence eliminating even more terms. One simple consequence of these relations that is important for the sequel is
\e{rhorho}{\bra{\rho} V^s_m \ket{\rho} \propto \delta_{s,2}\delta_{m,0}~.}

Thus, given the connection
\be\label{acon}
a=\Big(V^2_1 +{\alpha^2\over 4}V^2_{-1}+Q_s V^s_{-s+1}\Big)dw~,
\ee
our precise goal is to expand, perturbatively in $Q_s$, the chiral half of the entanglement entropy,
\be\label{entanglementform}
S_{\rm EE} = k_{CS} \log \Gc_\rho~,
\ee
with
\be
\Gc_\rho = \langle -\rho|e^{\Lambda}|\rho\rangle~, \quad \text{where} \quad \Lambda =  \Big(V^2_1 +{\alpha^2\over 4}V^2_{-1}\Big)w +Q_s V^s_{-s+1}w~.
\ee
The interval stretches from 0 to $w$. We expand the chiral Wilson line perturbatively as
\e{}{\Gc_\rho = \sum_{m=0}^{\infty} \Gc_\rho^{(m)}Q_s^m ~.}

\subsubsec{Zeroth order}

At zeroth order,
\be\label{lambda0}
\Lambda_0 \equiv \Lambda \big|_{Q_s=0} =  \Big(V^2_1 +{\alpha^2\over 4}V^2_{-1}\Big)w ~.
\ee
$e^{\Lambda_0}$ is in $SL(2)$, any element of which can be written as
\begin{align}
\label{ec}
e^{\Lambda_0} = e^{c_1 V^2_1} e^{\log c_0 V^2_0} e^{c_{-1} V^2_{-1}}~.
\end{align}
The coefficients $c_i$ are fixed by the group multiplication and are easily obtained by working in the defining representation. They are found to be
\begin{align}
\label{ed}
c_1 = \frac{2}{\alpha}\tan \frac{w\alpha}{2} ~,\quad c_0 = \cos^2 \frac{w\alpha}{2}~, \quad c_{-1} = {\a^2\over 4} c_1~.
\end{align}
Using the highest weight properties \eqr{hwlw}, we can then write the zeroth order matrix element, $\Gc_\rho^{(0)}$, as
\begin{align}
\label{ee}
\begin{split}
\Gc_\rho^{(0)} &= \braket{-\rho | e^{c_1 V^2_1} | \rho} c_0^{h_\rho}
\\
&= (c_0 c_1^2)^{h_\rho}\frac{\braket{-\rho | (V^2_1)^{2h_\rho} | \rho}}{(2h_\rho)!}
\\
&= \left(\frac{2}{\alpha}\sin \frac{w\alpha}{2}\right)^{2h_\rho}\frac{\braket{-\rho | (V^2_1)^{2h_\rho} | \rho}}{(2h_\rho)!}~,
\end{split}
\end{align}
where the second step is simply the statement that the height of the representation is $2h_\rho$.  (If $h_R$ is the $V^2_0$ eigenvalue of any highest weight state $\ket{\text{hw}_R}$, the height of the representation is $2h_R$.) The matrix element $\braket{-\rho | (V^2_1)^{2h_\rho} | \rho}$ can be calculated using standard techniques.  As is familiar from angular momentum theory, we have
\begin{align}
\label{ej}
\begin{split}
V^2_1\ket{h,m}
&= \sqrt{(h+m)(h-m+1)}\ket{h,m-1}
\end{split}
\end{align}
where $|h,m\rangle$ denotes a descendant state with $V^2_0=m$, belonging to the representation with highest weight $V^2_0=h$. From this we obtain
\begin{align}
\label{ek}
(V^2_1)^{2h_\rho}\ket{\rho} = (2h_\rho)! \ket{-\rho}
\end{align}
yielding for the zeroth order piece
\begin{align}
\label{el}
\Gc_\rho^{(0)}= \left(\frac{2}{\alpha}\sin \frac{w\alpha}{2}\right)^{2h_\rho}
\end{align}
and by \eqref{entanglementform}, the (chiral half of the) entanglement entropy
\begin{align}
\label{em}
\begin{split}
S_{\text{EE}} 
= \frac{c}{6}\log\left(\frac{2}{\alpha}\sin \frac{w\alpha}{2}\right)~.
\end{split}
\end{align}
This is the correct result. Namely, it is one half of $L/4G$, where $L$ is  the geodesic length computed in the metric
\begin{align}
ds^2 = d\rho^2 + e^{2\rho}\left| dw-{\alpha^2\over 4}e^{-2\rho}d\overline{w}\right|^2.
\end{align}

\subsubsec{Adding higher spin charge}\label{addinghsQ}
To study the case $Q_s \neq 0$, let us first develop some machinery to perform perturbation theory. The derivative of the exponential  can be written as
\begin{align}
\label{en}
\partial_{Q_s}\left(e^\Lambda\right) = \int_0^1 ds~ e^{s \Lambda}\partial_{Q_s} \Lambda~ e^{(1-s)\Lambda}~.
\end{align}
Differentiating again any number of times and setting $Q_s=0$, the exponentials are always SL(2) elements. For example the second order term is
\begin{align}
\label{eo}
\begin{split}
&\partial_{Q_s}^2\left(e^{\Lambda_0 + Q_s V^s_{-s+1}w}\right)\big|_{Q_s=0} \\
&~= \bigg[\int_0^1 ds_1 s_1\int_0^1 ds_2~ M_0(s_1 s_2)V^s_{-s+1}M_0(s_1(1-s_2))V^s_{-s+1}M_0(1-s_1) \\
&~+ \int_0^1 ds_1 (1-s_1)\int_0^1 ds_2~ M_0(s_1)V^s_{-s+1}M_0((1-s_1) s_2)V^s_{-s+1}M_0((1-s_1)(1-s_2))\bigg]w^2
\end{split}
\end{align}
where
\e{}{M_0(t) \equiv e^{t \Lambda_0}~.}
It is clear that the $Q_s^n$ term will have $n$ integrals and $n!$ terms, but all the terms are of the same type. Thus, the basic object we need to calculate is
\begin{align}
\label{ep}
{\cal J}_n(x_i) \equiv \braket{-\rho | M_0(x_1) V^s_{-s+1} M_0(x_2)\ldots V^s_{-s+1} M_0(x_{n+1}) | \rho}
\end{align}
for some parameters $x_i$.  In terms of ${\cal J}_n$, the second order perturbation of $\Gc_\rho$ reads\footnote{In writing out the arguments of ${\cal J}_n(x_i)$, we omit $x_{n+1} = 1-\sum_1^n x_i$.}
\es{j2}{\Gc_\rho^{(2)} = ~&w^2\int_0^1 ds_1\int_0^1 ds_2  \Big[s_1 \,{\cal J}_2\big(s_1s_2, s_1(1-s_2)\big)+
 (1-s_1) \,{\cal J}_2\big(s_1, s_2(1-s_1)\big)\Big]~.}

It is easy to see that  ${\cal J}_1$ vanishes for higher spin perturbations, $s>2$: equation \eqr{rhorho} holds, but $M_0(s_1) V^s_{-s+1} M_0(1-s_1)$ contains no $V^2_0$ contribution.\footnote{Likewise, the only way ${\cal J}_n$ can be non-zero, for any $n$, is if the normal ordering produces a $V^2_0$. It follows that ${\cal J}_n=0 $ for $n$ odd due to the grading of the SL(N) algebra.}  Therefore the entanglement entropy has no contribution linear in the higher spin charge. This matches reasoning from CFT from multiple vantage points. For example, recall that the Wilson line computes the semiclassical $W_N$ vacuum block. Decomposing the vacuum block into Virasoro blocks, the term linear in $Q_s$ corresponds to the exchange of the spin-$s$ current. (See Section \ref{conformalblocks} for more details.) This term is also linear in the higher spin charge $q_s$ carried by the light operators dual to the Wilson line probe. But the entanglement entropy probe has $q_{s>2}=0$, hence the entanglement entropy doesn't receive any contribution at linear order in $Q_s$.\footnote{Another, perhaps more direct, way to reach this conclusion is to note that the OPE coefficient between two twist operators and a spin-$s$ current vanishes.}

We focus now on the second order perturbation \eqr{j2}, and particularly the definition of the ${\cal J}_n$. Since each derivative only partitions the exponential, the $x_i$ simply comprise $n$ additive partitions of unity: $\sum_{i=1}^n x_i = 1$. As shown in Appendix \ref{smallchargedetails}, further use of SL(N) group theory puts this into a form more convenient for computation. First, introduce the following conjugated generator which obeys a first-order differential equation,
\e{}{V^s_m(t) \equiv e^{t\L_0} V^s_m e^{-t\L_0}~, \quad {d\over dt}V^s_m(t) = [\L_0,V^s_m(t)]~.}
Then we can write ${\cal J}_n$ as
\begin{align}
\label{ew}
\begin{split}
{\cal J}_n(x_i)&= \braket{\rho | \tilde{V}^s_{-s+1}(-1+x_1)\ldots \tilde{V}^s_{-s+1}(-1+\Sigma_1^n x_i) | \rho}\times \Gc_\rho^{(0)}
\end{split}
\end{align}
where
\e{}{\tilde{V}^s_{-s+1}(t) \equiv e^{-1/c_1 V^2_{-1}}\,V^s_{-s+1}(t) \,e^{1/c_1 V^2_{-1}}~.}
Recall \eqref{lambda0} and \eqref{ed} for the definitions of $\Lambda_0$ and $c_1$, respectively.

Because $(V^2_{-1})^{2h_\rho+1}=0$, $\tilde{V}^s_{-s+1}(t)$ is easily computed. Since the exponentials are SL(2) elements, $\tilde{V}^s_{-s+1}(t)$ can be written as a linear combination of spin-$s$ generators alone. Furthermore, using the Baker-Campbell-Hausdorff formula to calculate  $\tilde{V}^s_{-s+1}(t)$  involves only a finite number of commutators: in particular,
\e{adxy}{\text{ad}_{V^2_{-1}}^{m+s}\, V^s_{m} = 0~,\quad \text{where} \quad \text{ad}_X Y = [X,Y]~.}
The matrix elements ${\cal J}_n$ can then be obtained by normal ordering all the generators and using properties of the Weyl representation.

Let us carry out the above procedure for a spin-3 deformation. We begin by computing
\begin{align}
\label{ex}
\begin{split}
V^3_{-2}(t) &= w_{-2} V^3_{-2} + w_{-1} V^3_{-1} + w_0 V^3_0 + w_1 V^3_1 + w_2 V^3_2
\end{split}
\end{align}
with
\bea
\label{ezb}
w_{-2} &=& \cos^4 \tfrac{w \alpha t}{2},~w_{-1} = \frac{8}{\alpha}\cos^3 \tfrac{w \alpha t}{2}\sin \tfrac{w \alpha t}{2},~w_0 = \frac{6}{\alpha^2} \sin^2 w\alpha t,
\cr
w_1 &=& \frac{32}{\alpha^3} \cos \tfrac{w\alpha t}{2} \sin^3 \tfrac{w\alpha t}{2},~w_2 = \frac{16}{\alpha^4}\sin^4 \tfrac{w\alpha t}{2}~.
\eea
We then find
\begin{align}
\label{ey}
\begin{split}
\tilde{V}^3_{-2}(t) &= (w_{-2} + c_1^{-1} w_{-1} + c_1^{-2} w_0 + c_1^{-3} w_1 + c_1^{-4} w_2)V^3_{-2}
\\&+ (w_{-1} + 2 c_1^{-1} w_0 + 3 c_1^{-2} w_1 + 4 c_1^{-3} w_2) V^3_{-1}
\\&+ (w_0 +3 c_1^{-1} w_1 + 6 c_1^2 w_2)V^3_0 + (w_1 + 4 c_1^{-1} w_2)V^3_1 + w_2 V^3_2
\end{split}
\end{align}
and, plugging into \eqr{ew},
\begin{align}
\label{ez}
\begin{split}
{\cal J}_2(x_1,x_2) &=  \braket{\rho | (\tilde{w}_{-2} V^3_{-2} + \tilde{w}_{-1} V^3_{-1})(\tilde{w}'_1 V^3_1 + \tilde{w}'_2 V^3_2) | \rho}\times \Gc_\rho^{(0)}
\\
&= (N^2-4)h \left( \frac{4}{5}\tilde{w}_{-2}\tilde{w}'_2 -\frac{1}{10} \tilde{w}_{-1}\tilde{w}'_1 \right)\times \Gc_\rho^{(0)}
\end{split}
\end{align}
where the $\tilde{w}$ are functions of $(x_1-1)$ and $\tilde{w}'$ functions of $(x_1 + x_2 -1)$ that can be read off from \eqref{ezb}--\eqref{ey}. Note how few terms contribute to \eqr{ez} after using properties \eqref{ea}--\eqref{rhorho} of the Weyl representation. Finally, we integrate over $s_1$ and $s_2$ as in \eqr{j2} to obtain the second order result $\Gc_\rho^{(2)}$, and hence the chiral half of the entanglement entropy:
\begin{align}
\label{ezc}
\begin{split}
S_{\text{EE}} &= k_{CS} \log\left(\Gc_\rho^{(0)} + \Gc_\rho^{(2)} Q_3^2 + O(Q_3^4)\right)\\
&=\frac{c}{6}\log\left(\frac{2}{\alpha} \sin \frac{w\alpha}{2}\right) - \frac{(N^2-4)c}{80 \alpha^6}\csc^4 \tfrac{w\alpha}{2}\\ &\times\big[27 + (6w^2\alpha^2 -32)\cos w\alpha + 5\cos 2w\alpha +w\alpha(\sin 2w\alpha - 14 \sin w\alpha)\big]Q_3^2\\
&+O(Q_3^4)~.
\end{split}
\end{align}

\subsubsection{Entanglement in Vasiliev theory}
As explained in the introduction, the analytic continuation of \eqr{ezc} to \hsl, and hence to the higher spin sector of Vasiliev theory, is performed simply by replacing $N\rar\pm\l$. Therefore, we have arrived at a bulk derivation of the entanglement entropy in asymptotically AdS backgrounds of Vasiliev theory with perturbative spin-3 charge:
\begin{align}
\label{ezc2}
\begin{split}
&S_{\text{EE}} =\frac{c}{6}\log\left(\frac{2}{\alpha} \sin \frac{w\alpha}{2}\right)\\& - \frac{(\l^2-4)c}{80 \alpha^6}\csc^4 \left(\tfrac{w\alpha}{2}\right)\Big(27 + (6w^2\alpha^2 -32)\cos w\alpha + 5\cos 2w\alpha +w\alpha(\sin 2w\alpha - 14 \sin w\alpha)\Big)Q_3^2\\
&+O(Q_3^4)~.
\end{split}
\end{align}
This is one of our main results.

We now discuss the application to the \hsl\ higher spin black hole.  The computation above was based on the connection (\ref{acon}), which has $a_{\wb}=0$. The black hole solution has $a_w$ as in (\ref{acon}), but also a nonzero $a_{\wb}$ whose form is fixed by demanding trivial holonomy around the thermal circle of the black hole.     More precisely, $a_{\wb}$ is written in terms of the inverse temperature $\beta$ and spin-3 chemical potential $\mu$.  The holonomy condition relates these to the charges $\alpha$ and $Q_3$ \cite{Gutperle:2011kf}.  To the order we are working, the relations are\footnote{Note that $\alpha$ appearing here is not related to the parameter $\alpha$ appearing in \cite{Gutperle:2011kf}.}
\begin{align}
\label{ezd}
\alpha = \frac{2 \pi i}{\beta}\left(1 + \frac{40 \pi^2 \mu^2}{3 \beta^2}\right)~,~~ Q_3 = -\frac{8\pi^4}{3\beta^4}\sqrt{\frac{20}{N^2-4}} \mu~.
\end{align}
Suppose we compute the ``holomorphic" entanglement entropy for the higher spin black hole solution, as reviewed in appendix \ref{prescrips}.  This amounts to  ignoring $a_{\wb}$ but expressing the result in terms of $(\beta,\mu)$ using (\ref{ezd}).  This results in
\begin{align}
\label{eze}
\begin{split}
S_{\text{EE}} &= \frac{c}{6} \log \left(\frac{\beta}{\pi}\sinh x\right)
\\&+ \frac{c\pi^2}{36\beta^2}\mu^2\text{csch}^4 x\left(27 - 8(3x^2 + 4)\cosh 2x + 5\cosh 4x - 2x(\sinh 4x - 14 \sinh 2x)\right)\\
&+O(\mu^4)
\end{split}
\end{align}
with $x=\tfrac{\pi w}{\beta}$.

This is the result of the  holomorphic Wilson line, as defined in \cite{deBoer:2013vca}.   Remarkably, as shown in \cite{Datta:2014ska,Datta:2014uxa}, the result (\ref{eze}) precisely matches a CFT computation of entanglement entropy for a CFT deformed by a chemical potential for spin-3 charge.\footnote{Rather, it is exactly half of the CFT result in \cite{Datta:2014ska,Datta:2014uxa}, the other half coming from the anti-holomorphic piece and setting $\bar{\mu} = -\mu$.}   On the other hand, the agreement would not be present if we used the canonical Wilson line, which incorporate both $a_w$ and $a_{\wb}$ via $\Lambda= a_w w+ a_{\wb}\wb$.   It is possible that a modified CFT prescription yields agreement with the canonical Wilson line, and indeed this is understood to be the case for the thermal entropy of the black hole \cite{Compere:2013nba,deBoer:2014fra}.  The same procedure that shows how to go back and forth between the two versions of the  thermal entropy does not appear to work for the entanglement entropy \cite{Datta:2014ska,Datta:2014uxa}.   This deserves further investigation.

\subsection{Wilson lines for the $T=0$ higher spin black hole, and a match to CFT}

Another simple case we can consider is the following connection:
\be\label{chiraldefconnection}
a= V^2_1 dw -\mu V^3_{2}d\overline{w}~.
\ee
This describes the zero temperature limit of the higher spin black hole with spin-3 chemical potential $\mu$ fixed \cite{Gutperle:2011kf}. This connection was dubbed the ``chiral deformation'' in \cite{Ammon:2011nk}, as it can also be seen as dual to a deformation of the CFT vacuum by a constant source for a left-moving spin-3 current (cf. \eqr{source}). The chiral Wilson line $\Gc_R(\mu)$, written as a function of $\mu$, is then given by\footnote{Note that we perform our computation directly in the gauge presented in (\ref{chiraldefconnection}). Were we instead to transform this into highest weight gauge, which is always possible, we would not find the agreement with CFT that we discuss below.}
\begin{align}
\label{ga}
\Gc_{R}(\mu) &= \braket{-\text{hw}_R | e^{w V^2_1} e^{-\mu \bar{w} V^3_2} | \text{hw}_R}~.
\end{align}
Note that $[V^2_1, V^3_2] = 0$, which simplifies computations. This allows us to compute $\Gc_{R}(\mu)$ {\it non-perturbatively} in $\mu$ for simple enough representations, and to high orders in $\mu$ for the entanglement entropy.

We will then match our \hsl\ Wilson line result to a CFT computation at $\l=0$ using free fermions. The free fermion carries $W_{1+\infty}$ symmetry, which is closely related to the algebra $W_{\infty}[0]$, as we recall below. As the latter is the asymptotic symmetry of a bulk hs[0] theory, we may expect agreement between the Wilson line calculation and the CFT at this value of $\l$. Indeed, we verify agreement through $\mathcal{O}(\mu^4)$. This constitutes a highly non-trivial check of our Vasiliev Wilson line results.

\subsubsec{Warmup: $R=\Box$}
In the defining representation, $R=\Box$, the further relation $V^3_2 = (V^2_1)^2$ holds, which allows us to find $\Gc_{\Box}(\mu)$ exactly using SL(2) group theory alone. Expanding the two exponentials,
\begin{align}
\label{gb}
\Gc_{\Box}(\mu) = \sum_{m,n = 0}^\infty \frac{w^m (-\mu \bar{w})^n}{m!n!}\braket{N | (V^2_1)^{m+2n} | 1}
\end{align}
where $\ket{1}$ and $\ket{N}$ are the highest and lowest weights respectively of the defining representation. The matrix element vanishes unless the power of $V^2_1$ equals the height of the representation. The fact that $m$ is non-negative constrains the sum over $n$ to be over a finite set of values:
\es{gc}{\Gc_{\Box}(\mu) &= w^{2h} \sum_{n=0}^{\floor{h}}\frac{1}{(2h-2n)!n!}\left(\frac{-\mu \overline{w}}{w^2}\right)^n \braket{N | (V^2_1)^{2h} | 1}
\\
&= w^{2h} \sum_{n=0}^{\floor{h}}\frac{(2h)!}{(2h-2n)!n!}\left(\frac{-\mu \overline{w}}{w^2}\right)^n
\\
&=w^{2h}\left(\frac{-\mu \overline{w}}{w^2}\right)^{\floor{h}}\frac{(2h)!}{(2h-2\floor{h})!(\floor{h})!}~_2F_2\left(1,-\floor{h};\tfrac{1}{2}+h-\floor{h},1+h-\floor{h};\tfrac{w^2}{4\mu \overline{w}}\right)}
where $\floor{h}$ is the floor of ${N-1\over 2}$. The two cases where $h$ is an integer and a half integer correspond to $N$ being odd and even respectively. The Wilson line then becomes
\begin{align}
\label{gd}
\begin{split}
N \text{ odd }&:~ \Gc_{\Box}(\mu) = (-\mu \overline{w})^{\tfrac{N-1}{2}}\frac{(N-1)!}{((N-1)/2)!} ~_1F_1\left(\tfrac{1-N}{2};\tfrac{1}{2};\tfrac{w^2}{4\mu \overline{w}}\right)
\\
N \text{ even }&:~ \Gc_{\Box}(\mu) = w (-\mu \overline{w})^{\tfrac{N}{2}-1}\frac{(N-1)!}{(N/2-1)!} ~_1F_1\left(1-\tfrac{N}{2};\tfrac{3}{2};\tfrac{w^2}{4\mu \overline{w}}\right)~.
\end{split}
\end{align}

These results may be easily continued to \hsl\ to any fixed order in perturbation theory in $\mu$. From \eqref{gc}, the perturbative result is 
\begin{align}
\label{gf}
\begin{split}
\frac{\Gc_{\Box}(\mu)}{\Gc_{\Box}(0)} &= \sum_{n=0}^{\infty}\frac{1}{n!}\frac{\Gamma(N)}{\Gamma(N-2n)}\left(\frac{-\mu \overline{w}}{w^2}\right)^n
\\
&\rar \sum_{n=0}^{\infty} \frac{1}{n!}\frac{\Gamma(2n+1+\lambda)}{\Gamma(1+\lambda)}\left(\frac{-\mu \overline{w}}{w^2}\right)^n~.
\end{split}
\end{align}
The second line is the \hsl\ result.

We can compare this with calculations done in \cite{Gaberdiel:2013jca} directly in the context of Vasiliev theory. The bulk scalar of Vasiliev theory has the quantum numbers of the defining representation. Its propagator was computed as a solution to the higher spin wave equation in the $T=0$ black hole background of Vasiliev theory, from which its boundary two-point function was extracted. The result was (see equation 2.43 of \cite{Gaberdiel:2013jca})
\begin{align}
\label{gg}
\frac{\braket{\phi(w,\overline{w})\phi(0,0)}}{\braket{\phi(w,\overline{w})\phi(0,0)}_{\mu=0}} = \sum_{n=0}^\infty \frac{1}{n!}\frac{\Gamma(2n+1+\lambda)}{\Gamma(1+\lambda)}\left(\frac{\mu \overline{w}}{w^2}\right)^n~.
\end{align}
This matches our answer to any finite order in $\mu$ up to a sign, which is just the same difference of conventions explained in footnote 13. This constitutes another check on the analytic continuation. 

Non-perturbatively in $\mu$, the analytic continuation is not straightforward. The asymptotic expansion of the hypergeometric function $_1F_1$ is
\begin{align}
\label{ge}
\begin{split}
\frac{_1F_1(a;b;z)}{\Gamma(b)}\xrightarrow{|z| \text{ large}} \frac{e^{\pm i\pi a} z^{-a}}{\Gamma(b-a)}\left(\sum_{n=0}^{R-1} \frac{(a)_n(1+a-b)_n}{n!}(-z)^{-n} + \mathcal{O}(|z|^{-R})\right)
\\
+ \frac{e^z z^{a-b}}{\Gamma(a)}\left(\sum_{n=0}^{S-1}\frac{(b-a)_n(1-a)_n}{n!}z^{-n} + \mathcal{O}(|z|^{-S})\right)~.
\end{split}
\end{align}
The first sum in the expansion above reproduces the series we started with in \eqref{gc}. The second sum, on the other hand, doesn't contribute when $a$ is a negative integer, due to the gamma function in the denominator; this happens when $N$ is an integer. Without further analysis, we cannot unambiguously continue to \hsl\ non-perturbatively in $\mu$. See Section 4.4 of \cite{Campoleoni:2011hg} for analysis of a similar example that may be useful here.

\subsubsection{Entanglement entropy: $R=\rho$}

We now calculate the entanglement entropy in the $T=0$ higher spin black hole background \eqref{chiraldefconnection} from the Wilson line in the Weyl representation. This result is valid in the \hsl\ theory.

The Wilson line computation is a simpler version of the one in Section \ref{addinghsQ}, where now $c_1 = w,~ c_0 = 1$. We need the matrix element
\begin{align}
\label{gzza}
\Gc_\rho = \braket{ -\rho | e^{w V^2_1} e^{-\mu \bar{w} V^3_2} | \rho}
\end{align}
Using the techniques leading to \eqref{ew}, we can write
\begin{align}
\label{gzzb}
\begin{split}
\Gc_\rho &= w^{2h_\rho} \braket{ \rho | e^{-\tfrac{1}{w} V^2_1} e^{-\mu \bar{w} V^3_2} | \rho}
\\
&= w^{2h_\rho}\sum_{n=0}^{\infty} \frac{(-\mu \bar{w})^n}{n!} \braket{ \rho | \left(V^3_2 + \tfrac{4}{w}V^3_1 + \tfrac{6}{w^2} V^3_0 + \tfrac{4}{w^3}V^3_{-1} + \tfrac{1}{w^4} V^3_{-2}\right)^n | \rho}
\end{split}
\end{align}
The calculation to second order in $\mu$ is easily done by hand, but becomes tedious at higher orders. Using Mathematica, the chiral half of the entanglement entropy is found to be\footnote{We have now used the conventions of \cite{Ammon:2012wc} where $a_{\bar{w}} = -\mu N(\lambda) V^3_2$ and $N(\lambda) = \sqrt{\frac{20}{(\lambda^2-4)}}$}
\es{gzzc}{
S_{\text{EE}} &= k_{CS} \log \Gc_\rho
\\
& = \frac{c}{6}\log w - \frac{2c}{3}\frac{\bar{w}^2\mu^2}{w^4} - \frac{80c}{3}\frac{(2\lambda^2 - 17)}{(\lambda^2-4)}\frac{\bar{w}^4\mu^4}{w^8}-\frac{12800c}{63}\frac{(55\lambda^4 - 1259\lambda^2 + 6883)}{(\lambda^2-4)^2}\frac{\bar{w}^6\mu^6}{w^{12}}
\\
&~~ + \mathcal{O}(\mu^8)~.}
For a spacelike interval, $\bar w = w$.

We can try to compute the ``holomorphic Wilson line" as we did before for the higher spin black hole. To do this, simply ignore $a_{\overline{w}}$. For the chiral deformation, this leaves us with $\Lambda = w V^2_1$. Clearly, the ``holomorphic" entanglement entropy is then independent of $\mu$ and is given by just the vacuum result to any finite order in $\mu$. The canonical and holomorphic versions of thermal entropy were related by redefinition of charges (see \cite{Compere:2013nba,deBoer:2014fra} for details). However, this example illustrates that the same story cannot hold for the case of entanglement entropy.

\subsubsection{Matching to CFT at $\l=0$}
\label{CFTmatch}

Turning to the CFT side, we consider a theory of $N$ complex free fermions.   This theory has central charge $c=N$ and a higher spin symmetry algebra $W_{1+\infty}$, which contains currents of spins $s=1,2,\ldots \infty$.     The algebra $W_{1+\infty}$ reduces to $W_{\infty}[0]$ after removing the $U(1)$ current corresponding to fermion number.  In fact, in the computations that follow we are cavalier about the $U(1)$ current, but we comment on this at the end of this section.

We deform the free fermion action by a source for the spin-3 current $W(z)$ and then compute the entanglement entropy in conformal perturbation theory.  Since we are only interested here in  zero temperature we can consider the CFT on the plane.

A very similar calculation at finite temperature appears in \cite{Datta:2014ska}, whose notation we employ here. Consider a single interval running from $y_1$ to $y_2$, denoting the interval length as $\Delta\equiv y_2-y_1$. The R\'enyi entropy in conformal perturbation theory (with $\overline{\mu} = -\mu$) is given by (3.8 of \cite{Datta:2014ska})
\begin{align}
\label{gzzd}
S^{(n)}(\Delta) = \frac{1}{1-n}\log\frac{1}{Z^n}\prod_{a}\prod_{k=0}^{n-1}\bigg(\braket{\sigma^a_{k,n}(y_1,\bar{y}_1)\overline{\sigma}^a_{k,n}(y_2,\bar{y}_2)}_{\text{CFT}} +\tfrac{1}{2}\mu^2\int d^2z_1\int d^2z_2
\nonumber
\\
\braket{\sigma^a_{k,n}(y_1,\bar{y}_1)[W(z_1)+\overline{W}(\bar{z}_1)][W(z_2)+\overline{W}(\bar{z}_2)]\overline{\sigma}^a_{k,n}(y_2,\bar{y}_2)}_{\text{CFT}}+\ldots\bigg)
\end{align}
where $\sigma$ are the twist fields, $k$ being the replica index and $a$ counts the fermions.  Further, we can move to the bosonized language where we have an explicit representation of the twist operators (again see \cite{Datta:2014ska}). The twist fields and the spin-3 current in the bosonized form are given by
\begin{align}
\label{gzze}
\begin{split}
\sigma_{k,n}(z,\bar{z}) &=~ :\!\prod_{a}e^{\frac{ik}{n}\left(\phi_{a,k}(z)-\bar{\phi}_{a,k}(\bar{z})\right)}\!:
\\
\overline{\sigma}_{k,n}(z,\bar{z}) &=~ :\!\prod_{a}e^{-\frac{ik}{n}\left(\phi_{a,k}(z)-\bar{\phi}_{a,k}(\bar{z})\right)}\!:
\\
W &= -\frac{\sqrt{5}}{6 \pi}\sum_a :\!(\partial \phi_a)^3\!:
\end{split}
\end{align}
In the bosonized language, everything is determined by the $\phi\phi$ OPE. Some useful relations  are
\begin{align}
\label{gzzf}
\begin{split}
&\sigma^a_{k,n}(y_1,\bar{y}_1)\overline{\sigma}^a_{k,n}(y_2,\bar{y}_2) = \frac{:\!\sigma^a_{k,n}(y_1,\bar{y}_1)\overline{\sigma}^a_{k,n}(y_2,\bar{y}_2)\!:}{\Delta^{2k^2 N/n^2}}
\\
&\partial \phi_a(z):\!\sigma^a_{k,n}(y_1,\bar{y}_1)\overline{\sigma}^a_{k,n}(y_2,\bar{y}_2)\!: \sim \left(\frac{ik}{n}\right)\frac{\Delta}{(z-\Delta)} :\!\sigma^a_{k,n}(y_1,\bar{y}_1)\overline{\sigma}^a_{k,n}(y_2,\bar{y}_2)\!:
\\
&\partial \phi_a(z_1) \partial \phi_b(z_2) \sim -{1\over (z_1-z_2)^2}
\end{split}
\end{align}
 To facilitate calculations on a computer, we first replace $:\!(\partial \phi)^3\!:$ by $e^{\varepsilon \partial \phi}$ and note that
\begin{align}
\label{gzzg}
:\!(\partial \phi)^3\!:~ = \frac{\partial^3}{\partial \varepsilon^3}:\!e^{\varepsilon \partial \phi}\!: \bigg|_{\varepsilon = 0}~.
\end{align}
The correlator then involves only exponential of operators, and this can be automated easily. All that's left is to perform the integrals in \eqref{gzzd}. At zero temperature, all the integrands have a very simple form: inverse of a polynomial in $z_i$. To integrate on the complex plane, we use the prescription of \cite{Hijano:2014sqa}
\begin{align}
\label{gzzh}
\frac{1}{(z-z_1)^m_1(z-z_2)^m_2} = \partial_{\bar{z}}\left(\frac{\bar{z}}{(z-z_1)^m_1(z-z_2)^m_2}\right)- \bar{z}\partial_{\bar{z}}\left(\frac{1}{(z-z_1)^m_1(z-z_2)^m_2}\right)~.
\end{align}
We also use the relation
\begin{align}
\label{gzzi}
\partial_{\bar{z}}\left(\frac{1}{(z-z_0)^{m+1}}\right)=2\pi\frac{(-1)^m}{m!}\partial_z^m\delta^{(2)}(z-z_0,\bar{z}-\bar{z_0})~.
\end{align}
Repeating the procedure for multiple integrals, we find for the entanglement entropy,
\begin{align}
\label{gzzj}
S_{\text{EE}} = \frac{c}{3}\log\Delta -\frac{4c}{3}\frac{\mu^2}{\Delta^2} - \frac{680c}{3}\frac{\mu^4}{\Delta^4} + \mathcal{O}(\mu^6)~.
\end{align}
Recalling that $\Delta$ is the interval length, which we denoted $w$ in the previous subsection, the CFT result matches the bulk calculation \eqref{gzzc} at $\lambda = 0$, up to an overall factor of 2. This factor is expected, as we only included the contribution from the chiral half of the Wilson line.

This match between bulk and CFT calculations is the entanglement analog of the results of \cite{Kraus:2011ds}, which included a match between thermal partition functions of the free fermion CFT and the bulk \hsl\ black hole.

A couple of comments are in order. As mentioned earlier, all our calculations are in the free fermion CFT which has $W_{1+\infty}$ symmetry. A more faithful calculation would be to directly use the currents of \cite{Gaberdiel:2013jpa}, which have $W_\infty[0]$ symmetry, in \eqref{gzzd}. However, the form of the twist operators are not known in the coset CFT. In \cite{Datta:2014uxa}, the authors sketch an argument for the agreement found between entanglement entropies in the two CFTs at order $\mu^2$. Our result hints at an extension of this to higher orders in $\mu$, but such a generalization eludes us.

There are also various different prescriptions for carrying out perturbation theory, as noted in appendix \ref{prescrips}.
Here we have calculated the entanglement entropy in `holomorphic' perturbation theory, as in \cite{Datta:2014ska}. However, we find that the results match up with the `canonical' Wilson line in the bulk. We believe this might have to do with our choice of prescription to perform the integrals in the CFT. It would be nice to understand this relation better.

\subsubsection{Aside: higher spin charges of arbitrary representations}\label{probe charges}

As follows from the discussion below \eqr{adxy}, the piece of the matrix element \eqr{ga} linear in $\mu$ is proportional to the spin-3 charge carried by the Wilson line probe. In the case $R=\square$, for example, the linear term of \eqr{gf} is
\begin{align}
\label{gza}
\Gc_{\Box}(\mu) =  \left( 1 - (1-N)(2-N) \frac{\mu \bar{w}}{w^2} + \mathcal{O}(\mu^2)\right)\Gc_{\Box}(0)~.
\end{align}
With the continuation $N \to -\lambda$ we infer that the spin-3 charge in the defining representation must be proportional to $(1 + \lambda)(2 + \lambda)$ which agrees with the CFT result in \cite{Gaberdiel:2013jca}. In fact, we can use this technique to calculate the charge carried by the probe in any representation: simply compute the piece of
\begin{align}
\label{apa0}
\Gc_{R}(\mu) = \braket{-\text{hw}_R | e^{w V^2_1 - \mu \overline{w} V^s_{s-1}} | \text{hw}_R}
\end{align}
that is linear in $\mu$, for general $R$.

It is sufficient to find the charges in the antisymmetric tensor representations -- call them $q_{s,k}$ -- since all other representations can be constructed out of tensor products of these:
\e{gzba}{q_{s,R} = \sum_k q_{s,k} d_k}
where $d_k$ are the Dynkin labels of the representation $R$.

As in Section \ref{closed}, the matrix elements in the ${\bf {\rm \bf asym}_k}$ representation are related to the matrix elements in the defining representation,
\begin{align}
\label{gzb}
G_k &= \det \tilde{A}~,\quad \tilde{A}_{ij}=\braket{N+1-i | e^{w V^2_1} e^{-\mu\bar{w} V^3_2} | j}~\quad i, j=1, 2, \ldots k~.
\end{align}
We can calculate these matrix elements using the relation $V^3_2 = (V^2_1)^2$, which only holds in the defining representation, and to first order in $\mu$ obtain
\begin{align}
\label{gzc}
\tilde{A}_{i,j} = \frac{w^{N+1-j-i}}{(N+1-j-i)!}\sqrt{\frac{N-j)!}{(j-1)!}\frac{(N-i)!}{(i-1)!}}\left(1 - (N+1-j-i)(N-j-i)\frac{\mu\bar{w}}{w^2} +\mathcal{O}(\mu^2)\right)
\end{align}

To calculate these determinants $G_k$, we perform column operations to reduce the determinant of the $k\times k$ matrix to that of a $(k-1)\times (k-1)$ matrix and so on. The calculations are fairly straightforward but tedious and are relegated to appendix \ref{probe charge}. Computing the Wilson line and reading off the term linear in the spin-$s$ chemical potential gives the spin-$s$ charges in the ${\bf {\rm \bf asym}_k}$ representation, up to normalization:
\begin{align}
\label{gzn}
q_{s,k} \propto \frac{\Gamma(N-k+1)}{\Gamma(N-s+1)}\sum_{r=0}^{s-2}(-1)^r\frac{(s-1-r)_k(N-k+1)_r(N-s+2-k+r)_{k-1-r}}{(s-1)\Gamma(k-r)\Gamma(r+1)}
\end{align}
where $(a)_n = a(a+1)\ldots (a+n-1)$ is the ascending Pochhammer symbol. A simple check of our result is that the higher spin charge must vanish if we choose $N < s$, which is indeed manifest here.

The proportionality constant can be fixed using known results for the higher spin charges in the defining representation (see e.g. equation 5.18 of \cite{Ammon:2011ua}). After doing so, we can continue these to \hsl\ by taking $N\to -\lambda$, noting that all $N$-dependence is polynomial. The final result for the spin-$s$ charge carried by the $\asym_k$ representations of \hsl\ is
\begin{align}
\label{gzna}
q_{s,k} = \frac{\Gamma(s)^2}{\Gamma(2s-1)}\frac{\Gamma(\lambda+s)}{\Gamma(\lambda+k)}\sum_{r=0}^{s-2}(-1)^r \frac{(s-1-r)_k(\lambda+k-r)_r(\lambda+s)_{k-1-r}}{(s-1)\Gamma(k-r)\Gamma(r+1)}~.
\end{align}

The charges for low-lying values of $s$ and $k$ are given in equation \eqr{gzo}. They agree with known results for $s=2,3,4$ in the $k=1,2$ representations in \cite{Gaberdiel:2013jca}. Another useful check is that the Weyl representation should carry no higher spin charge. We have checked for several values of $s>2$ that $q_{s,\rho}$ indeed vanishes. 

\subsection{Short interval expansion of entanglement entropy}\label{shortint}

Another useful application is to calculate the entanglement entropy in a short interval expansion. Interestingly, this can be done non-perturbatively in the charges; the small parameter is instead the interval size. On a technical level, this calculation is similar to, but simpler than, the small charge expansion.

\subsubsec{CFT expectations}

Let us get oriented using arguments from CFT \cite{Calabrese:2010he, Headrick:2010zt, Hartman:2013mia, Chen:2013dxa, Perlmutter:2013paa}. The single interval entanglement in a higher spin-excited state, call it $|\Psi_{\rm HS}\rangle$, is computed from a twist field two-point function in that state. This may be done by using the replica trick to evaluate the R\'enyi entropy $S_n$ at arbitrary R\'enyi index $n$, and then taking $n\rar 1$ to recover the entanglement entropy:
\e{}{S_{EE} = \lim_{n\rar 1} {1\over 1-n}\log \langle\Psi_{\rm HS}|\Phi_+(w)\Phi_-(0)|\Psi_{\rm HS}\rangle_{{\cal C}^n/\mathbb{Z}_n}~.}
${\cal C}$ is the original CFT, and the correlator is evaluated in its $n$-fold cyclic orbifold.

Developing a short interval expansion means that we use the OPE between the twist fields \cite{Calabrese:2010he}. Each power of the interval length that appears is equal to the conformal dimension of an operator in the OPE. Finding the leading contribution, therefore, from the existence of a spin-$s$ current in the CFT boils down to finding the primary operator of lowest conformal dimension in the ${\cal C}^n/\mathbb{Z}_n$ orbifold theory that involves the current and that appears in the twist field OPE. As explained in \cite{Perlmutter:2013paa}, this operator is comprised of two spin-$s$ currents living on different sheets, which contributes to the OPE as $w^{2s}$ relative to the identity. So the leading contribution to the short interval expansion of the entanglement entropy from a spin-$s$ background charge will look like
\begin{align}
\label{ezm}
S_{\text{EE}}(Q_s) - S_{\text{EE}}(0) = c f_s(N)Q_s^2 w^{2s} + \mathcal{O}(w^{2s+2})
\end{align}
for some function $f_s(N)$ polynomial in $N$. Note that mixing of $Q_s$ with other charges occurs only at higher orders. 

The polynomial $f_s(N)$ is determined in CFT by the one-point function of the aforementioned operator in the excited state $|\Psi_{\rm HS}\rangle$. We will find this polynomial for $s=3$ using the bulk Wilson line instead. Afterwards, we will give a simple argument for the $N$-dependence of $f_s(N)$ for any $s$.

\subsubsec{Wilson line calculation}

Again, we consider the asymptotically AdS background connection \eqr{acon},
\e{}{a_w = V^2_1 + \sum_{s=2}^{\infty}Q_s V^s_{-s+1}~.} 
We are interested specifically in the leading order effect of spin-$s$ charges of $s>2$ in the short interval expansion, hence we can safely turn off the spin-2 charge in our connection.\footnote{In the absence of $s>2$ charges, the spin-2 charge is trivial to incorporate into the entanglement entropy, for {\it arbitrary} interval size: in that case, $S_{EE}$ is given in \eqr{em}. In the presence of a spin-$s>2$ charge, the spin-2 and spin-$s$ charges will only mix at $O(w^{2s+2})$ in the short interval expansion; this is clear from the CFT argument using the twist OPE.} Henceforth we turn on only $Q_3$, which captures the leading correction due to higher spin charge. Our goal, as before, is to calculate the Wilson line in the Weyl representation,
\begin{align}
\label{ezf}
\Gc_\rho(Q_3) = \braket{-\rho | e^{w(V^2_1 + Q_3 V^3_{-2})} | \rho~.}
\end{align}
Expanding \eqr{ezf} in small $w$,
\e{}{\Gc_\rho(Q_3) = \sum_{n=0}^{\infty}{w^n\over n!} \braket{-\rho | (V^2_1 + Q_3 V^3_{-2})^n | \rho}~.}
It is clear from our previous manipulations that the leading term appears at $n=2h_{\rho}$, and the leading correction due to $Q_3$ appears at $n=2h_{\rho}+6$. The latter term is proportional to $Q_3^2$ -- in agreement with our CFT expectations -- despite being non-perturbative in $Q_3$. This means that we can pretend that we are working to quadratic order in small $Q_3$. But we already derived the result for small $Q_3$, and for {\it arbitrary} $w$, in Section \ref{smallcharge}! Therefore we may read off the result from the $w\ll1$ expansion of \eqr{ezc}. In the parameterization \eqr{ezm}, one finds
\e{f3}{f_3(N) = -{N^2-4\over 42000}}
which is our desired result. (Note that while \eqr{ezc} depends on spin-2 charge $\a$, \eqr{f3} does not, as explained above.) This can be easily checked for any fixed $N$ using our explicit Wilson line of Section \eqr{closed}. Lest this appear too indirect, we give a direct calculation of $f_3(N)$ from the matrix element \eqr{ezf}, without using our previous small charge results, in Appendix \ref{shortinterval}. 

As with our other computations, the analytic continuation to the \hsl\ theory is manifest: simply replace $N\rar\pm\l$ in \eqr{f3}. This constitutes a computation in the \hsl\ theory that is non-perturbative in higher spin charge.

Generalizing to arbitrary $s$, the $N$-dependence of the $O(Q_s^2)$ term can be fixed by the following argument. It is clear that $f_s(N)$ is a polynomial in $N$, determined as it is by SL(N) structure constants. Indeed, the polynomial is fixed by a single structure constant: from the above calculations and those in Appendix \ref{shortinterval}, the $O(Q_s^2)$ term is produced by commutators of the form $[V^s_m, V^s_{-m}]$ sitting inside the matrix element. The only term that survives the commutator is the spin-2 piece.
Looking at the structure constants in Appendix \ref{slngroup}, specifically $g^{ss}_{2s-2}(m,-m; N)$, one finds
\e{}{\bra{\rho} [V^s_m, V^s_{-m}]\ket{\rho} \propto h_{\rho}(N^2-4)(N^2-9)\cdots (N^2-(s-1)^2)~.}
This is the full $N$-dependence of the $O(Q_s^2)$ result, which determines $f_s(N)$ up to an overall, probably $s$-dependent, normalization factor. Continuation to \hsl\ is trivial.

\subsection{Virasoro conformal blocks from $W_N$ conformal blocks}\label{conformalblocks}

As reviewed in Section \ref{vasilievwilsonlines}, the Wilson line computes the semiclassical $W_N$ vacuum block.   The states spanning the $W_N$ vacuum module can be reorganized in representations of the Virasoro algebra. Therefore, the $W_N$ vacuum block can be written as a sum over vacuum and non-vacuum Virasoro blocks, with the same external states. In this section we show how the perturbative expansion of the Wilson line can be used to extract the semiclassical Virasoro block for pairwise identical external operators, and arbitrary internal operator dimension.

The philosophy is simple: allow the chiral Wilson line probe to carry spin-$s$ charge, and extract the part of the Wilson line linear in this charge. This must be proportional to the Virasoro block for the spin-$s$ current exchange. See Figure \ref{fig}. Because the Virasoro blocks depend rationally on the internal operator dimension, we can analytically continue $s$ to an arbitrary real number, yielding the general result.\footnote{Similar ideas were recently employed in \cite{Hijano:2015qja}.}
\vs

We first quote the formula for the Virasoro block in the semiclassical limit. We consider a four-point function with external operators of equal chiral dimensions $H_{1}=H_2$ and $h_{1}=h_2$, and with an exchanged primary of dimension $h_p=s$.  $H_{1,2}$ are dimensions of heavy operators, with the rest being ``light'' operators. The semiclassical limit is the same one given in \eqr{sclim}: taking $c\rt \infty$, holding fixed $H_1/c$ and $h_{1,p}/c\ll 1$.  The contribution to the four-point function in this limit was computed in \cite{Fitzpatrick:2015zha}.\footnote{As established there, the semiclassical limit of the Virasoro block for pairwise identical external dimensions is actually the same as the heavy-light limit, in which $h_{1,p}$ are held fixed at large $c$.} Writing the result on the cylinder, as in equation 2.13 of \cite{Hijano:2015qja}, gives the following prediction for the Wilson line to linear order in the spin-$s$ charge
\be\label{cfthyp}
\Gc(Q_s)  = \left(\sin{\alpha w\over 2}\right)^{-2h_1} \left(1+ C_s Q_s q_s  (1-e^{i\alpha w})^{s}{_2}F_1\left(s,s,2s;1-e^{i\alpha w}\right) + \ldots          \right)
\ee
where $C_s$ is some $w$-independent constant.

We can extract this term using the same small charge expansion of Section \ref{smallcharge}, only we need to allow the Wilson line probe to carry some spin-$s$ charge. The simplest choice is to use the defining representation. Denoting the highest weight state by $\ket{1}$ and the lowest weight state by $\ket{N}$, we have
\begin{align}
\label{ezza}
V^2_{-1}\ket{1} = 0 = V^2_1\ket{N}~,~~ V^2_0\ket{1} = \frac{(N-1)}{2}\ket{1}~,~~ V^2_0\ket{N} = -\frac{(N-1)}{2}\ket{N}~.
\end{align}
A convenient property of the defining representation is the simple relation $V^s_{-s+1} = (V^2_{-1})^{s-1}$.

As in Section \ref{smallcharge}, we are interested in the object
\begin{align}
\label{ezzb}
\begin{split}
{\cal J}_1(t) &= \braket{N | e^{\Lambda_0}V^s_{-s+1}(t) | 1}
\\
&= \braket{N | e^{\Lambda_0} (V^2_{-1}(t))^{s-1} | 1}
\\
&= (c_0 c_1^2)^h \braket{1 |(\tilde{V}^2_{-1}(t))^{s-1} | 1}
\end{split}
\end{align}
where all symbols have been defined earlier in Section \ref{addinghsQ}. This gives the first order Wilson line as
\be
\Gc_{\Box}(Q_s) = \Gc_{\Box}(0)\left(1+ A_s(w)Q_s+\ldots\right)
\ee
with
\e{ezze2}{A_s(w)  \equiv w \int_0^1 dt~{\cal J}_1(t-1)~.}
Looking back at (\ref{cfthyp}), we are aiming to show
\be\label{Aform}
A_s(w) \propto (1-e^{i\alpha w})^{s}{_2}F_1\left(s,s,2s;1-e^{i\alpha w}\right)   ~. \ee
We now note that the hypergeometric functions ${_2}F_1(s,s;2s;z)$ with $s\in \mathbb{Z}^+$ obey a recursion relation\footnote{The relevant relation is ${_2}F_1(a,b;2b;z) =2^{2b} \pi^{-1/2}\frac{\Gamma(b+1/2)}{\Gamma(2b-a)}z^{-b}(1-z)^{1/2(b-a)} P^{b-a}_{b-1}(\frac{2}{z}-1)$, and the Legendre functions obey a recursion relation.}
\begin{align}
\label{ezzf}
\begin{split}
F(s,s;2s;z) = \frac{2(2s-1)(2s-3)}{(s-1)^2}\bigg(\left(\frac{2-z}{z^2}\right)F(s-1,s-1;2(s-1);z)
\\-\frac{2}{z^2} F(s-2,s-2;2(s-2);z)\bigg)~.
\end{split}
\end{align}
Our basic strategy  is to show that  $A_s(w)$ obeys a corresponding recursion relation. This is done in Appendix \ref{wnvir}, and the result is that we arrive at the desired relation (\ref{Aform}).

\section{Relation to $W_N$ Minimal Model Holography}\label{wnsec}
Here we explain how the foregoing relates to the holographic duality between Vasiliev theory and the semiclassical limit of the $W_N$ minimal models.

The salient aspects of that duality are as follows. On the CFT side, one begins with the $W_N$ coset $SU(N)_k \times SU(N)_1/SU(N)_{k+1}$, where $N\in\mathbb{Z}$, and takes the semiclassical limit $k \rar -N-1$. The central charge grows linearly as $c \sim 1/(k+N+1)$, and the theory behaves like a vector model with $W_N$ symmetry. The bulk dual is Vasiliev's higher spin gravity, which contains a free parameter $\lambda$. Among other roles, $\lambda$ appears in the asymptotic symmetry of the AdS$_3$ vacuum, namely, the classical $W_{\infty}[\l]$ algebra. In the duality with the semiclassical limit of the coset, one takes $\lambda=\pm N$, which yields a symmetry algebra $W_{\infty}[\pm N] \cong W_N$. This contrasts with the original conjecture of Gaberdiel and Gopakumar, in which one takes the ``'t Hooft limit'' of the coset, of large $N,k$ with $\lambda=N/(N+k)$ fixed, where $\lambda$ is identified with the $\l$ of the Vasiliev theory. 

The semiclassical coset theory is non-unitary, which may be attributed to the fact that $k<0$. Nevertheless, it is a sparse CFT which obeys large $c$ factorization. In the limit, all light scalar operators $\Oc_L$ are specified by Young diagrams of SU(N): $\Oc_L = (\L_+,0)$, where $\L_+$ denotes a Young diagram. All $\Oc_L$ have negative conformal dimensions in this limit, hence the non-unitarity. (In view of their negative dimensions, ``light'' means $|h|\sim O(c^0)$.) Convincing evidence has been presented that the duality makes sense perturbatively \cite{Perlmutter:2012ds,Hijano:2013fja}, including the match of four-point functions described in Section \ref{vasilievwilsonlines}.

Returning to the realm of Wilson lines, we may now ask what they tell us about the minimal models. The semiclassical coset duality provides an operator realization of our general results: the CFT furnishes a tower of operators whose quantum numbers are determined by the same Young tableaux that specify the probe charges of the Wilson line. This gives an operator application of the representation theory for the finite-dimensional representations, where we identify $R=\L_+$. We explained in Section \ref{conformalblocks} that the Wilson line is essentially identical to the heavy-light four-point function in Vasiliev theory at $\l=\pm N$. So invoking the holographic duality, we land on a nice interpretation of the Wilson line as computing semiclassical coset correlators. As explained in Section \ref{vasilievwilsonlines}, it follows that these are dominated by the $W_N$ vacuum block.

We may also interpret the computation in Section \ref{probe charges}, of all higher spin charges $q_s$ of arbitrary representations $R$, as the charges of the semiclassical coset operators $(R,0)$. In fact, these are also the charges of the operators $(R^T,0)$ of the coset in the {\it `t Hooft limit} as well, upon continuing $N \rar \pm \lambda$. This follows from the conjecture in Section 6.2.2 and Appendix C of \cite{Gaberdiel:2011zw}:
\e{scth}{q_s^{\rm `t~Hooft}(R^T) = q_s^{\rm semiclassical}(R)~.}
Our result \eqref{gzna} gives the right-hand side of this equality. This is a new piece of data in the `t Hooft limit: as explained in \cite{Gaberdiel:2011zw}, formulas for higher spin charges of spin $s>2$ of coset operators $(R,0)$ in the so-called primary basis \eqr{hlcharges} are not known, and hence their `t Hooft limit cannot be taken. We have used analytic continuation from the semiclassical limit to derive these charges directly in the `t Hooft limit, assuming \eqr{scth}.

\section{Discussion}\label{disc}

Let us briefly summarize. The focus of this work was on developing efficient methods for computing
Wilson line observables in higher spin theories.   Perhaps the main
advance is that we were able to obtain results in SL(N) theory for
arbitrary N; this, together with some information about their analytic
structure, allowed us to apply the analytic continuation $N\rightarrow
-\lambda$ to obtain results in the \hsl\ Chern-Simons theory. The latter governs the higher spin sector of the 3D Vasiliev theory that appears in the Gaberdiel-Gopakumar
duality conjecture to $W_{\infty}[\lambda]$ coset CFTs.  Via several explicit calculations, we showed how such results match up with
results obtained from CFT perturbation theory in higher spin charges and
potentials. Taking the charges of the Wilson line to be those of a CFT twist operator, with
a nonzero energy but vanishing higher spin charge, leads to an object that computes entanglement entropy. For general charges, the Wilson line computes the
semiclassical $W_N$ vacuum block in the limit \eqr{sclim}. We provided another confirmation of this correspondence between Wilson lines and conformal blocks,
by verifying that the $W_N$ vacuum block so obtained decomposes as
expected into Virasoro blocks upon expanding to first order in the charges.

A logical question is whether some
further extension would allow for the computation of non-vacuum
blocks of the $W_N$ algebra.   Actually, this statement requires
refinement because non-vacuum $W_N$ blocks cannot be defined by direct
analogy to the Virasoro case.  In the determination of Virasoro
blocks, one uses that the three-point function of primaries fully
determines the three-point function of descendants, but this statement
no longer holds when the word Virasoro is replaced by $W_N$: instead, a (generically infinite) set of new parameters appears in the $W_N$ case \cite{Bowcock:1993wq}, and
the definition of non-vacuum $W_N$ blocks appears ambiguous (though see \cite{Gavrylenko:2015wla}). Keeping this
complication in mind, it is natural to expect that semiclassical
non-vacuum $W_N$ blocks have a holographic representation in terms of
Wilson lines with junctions; this would be the analog of the worldline
picture for non-vacuum Virasoro blocks \cite{Hijano:2015qja}. It would be very
interesting to establish such a relation, or to motivate a physical prescription for defining the non-vacuum $W_N$ blocks by using nice properties of Wilson line junctions.

It is natural to wonder if there is a formulation of the Wilson line
that would apply directly in the \hsl\ theory, obviating the
need to analytically continue from SL(N).   It is easy to write down
formal expressions for such a Wilson line, and indeed, we have done so in this paper: the expression \eqr{wilson2} is formally valid, where the sum in $P_0$ now runs to infinity. But to actually evaluate them requires a better understanding of what it means to exponentiate
elements of \hsl\ that lie outside an SL(2) subalgebra.

Actually, this is one of several related issues that have obstructed a non-perturbative understanding of Vasiliev theory and \hsl\ Chern-Simons theory in general. To gain some perspective, we put entanglement aside and recall the earlier calculations of the \hsl\ black hole partition function, which is a conceptually simpler problem. The calculations of \hsl\ higher spin black hole thermodynamics were performed both in the bulk \cite{Kraus:2011ds} and in CFT \cite{Gaberdiel:2012yb}, but only perturbatively in the spin-3 chemical potential $\mu$ on boh sides. In the bulk, one cannot even construct the smooth solution non-perturbatively, much less compute its partition function: the smoothness prescription involves computing the holonomies of the connections around the thermal circle, but these require exponentiating to the group HS[$\l$], which we cannot yet do in general. In the CFT, the problem is to find the modular properties of the grand canonical partition function $Z= \Tr(q^{L_0} y^{W_0})$ under SL(2,$\mathbb{Z}$) transformations. Unlike for the case of U(1) charged black holes and weak Jacobi forms, these are not known. See \cite{ Monnier:2014tfa, Li:2013rsa, Iles:2013jha,Iles:2014gra} for some progress on these questions.

Returning to the arena of entanglement, we see that the group HS[$\l$] would be needed here too. The CFT dual of the \hsl\ Wilson line is the semiclassical $W_{\infty}[\l]$ vacuum block, and as one should expect, there is a clear CFT dual of the non-perturbative hurdles in the bulk. To compute the block directly in CFT, one resorts to a monodromy prescription. For $W_N$, this method requires solution of an order-$N$ differential equation; but for $W_{\infty}[\l]$, the differential equation is of infinite order. Perhaps studying its structure would shed light on the non-perturbative \hsl\ Wilson line problem. While we have taken useful first steps in the present work, fuller answers to these questions would potentially give a non-perturbative definition of the ``higher spin geometry'' of Vasiliev theory and point the way toward a deeper understanding of string theory.


\vspace{.3in}

\noindent
{ \bf \Large Acknowledgments}

\vspace{.1in}
We are grateful to Alejandra Castro, Eliot Hijano and Eva Llabr\'es for helpful discussions. P.K. is supported in part by NSF grant PHY-1313986. E.P. wishes to thank the Perimeter Institute for Theoretical Physics, the Kavli Institute for Theoretical Physics, Strings 2015 and the Simons Center for Geometry and Physics for hospitality during this project. This research was supported in part by the National Science Foundation under Grant No. NSF PHY11-25915. Research at Perimeter Institute is supported by the Government of Canada through Industry Canada and by the Province of Ontario through the Ministry of Economic Development and Innovation. E.P. is supported by the Department of Energy under Grant No. DE-FG02-91ER40671.

\appendix
\section{SL(N) group theory}\label{slngroup}
\label{group}

For convenience we collect some relevant facts concerning SL(N) group theory, all of which are completely standard.  In the Cartan-Weyl basis the Lie algebra is
\bea[H^i,H^j]&=&0\cr
[H^i,E^\alpha]& =&\alpha^i E^\alpha\cr
[E^\alpha,E^\beta]& =& \left\{\begin{array}{cc} \alpha\cdot H &   {\rm if}~ \alpha+\beta=0 \cr
0& {\rm otherwise}\end{array} \right.
\eea
Here $i = 1,2, \ldots , N-1$.
The roots obey $\alpha\cdot \alpha =2$.      Dot products are taken with respect to the Killing metric, which we take to be defined by the trace in the defining representation,
\be \Tr_\Box (H^i H^j) =\delta_{ij}~,
\ee
so that the Killing metric in this basis is $\eta_{ij}=\delta_{ij}$, hence $\alpha \cdot \alpha = \sum_i \alpha^i\alpha^i$.

We denote by $|\lambda\rangle$ the state associated with weight vector $\lambda$:  $H^i |\lambda\rangle=\lambda^i|\lambda\rangle$. Associated to any weight $\lambda$ is a dual element of the Cartan subalgebra, $\lambda \cdot H$, and vice versa.  The roots $\alpha$ are the weights of the adjoint representation.    We denote the $N-1$ simple roots as $\alpha^{(i)}$.  Recall that a simple root is root that cannot be written as a sum of two positive roots.  Positivity is defined by choosing some basis for the root space, expanding in this basis, and requiring that the first (say) nonzero expansion component is positive.   The dot product of the simple roots defines the Cartan matrix,
\be
A_{ij} = \alpha^{(i)} \cdot \alpha^{(j)}~.
\ee
For SL(N) the nonzero entries of the Cartan matrix are
\be
A_{ii}=2~,\quad A_{i,i\pm 1}=-1~.
\ee

We  define $N-1$ fundamental weights $\omega^{(i)}$ to obey
\be
 \alpha^{(i)} \cdot \omega^{(j)}= \delta_{ij}~.
\ee
They can be expressed in terms of the simple roots as
\be
\omega^{(i)} = \sum_j A^{-1}_{ij}\alpha^{(j)}~,
\ee
and so
\be
\omega^{(i)} \cdot \omega^{(j)} = A^{-1}_{ij}~.
\ee
For SL(N),
\be
 A^{-1}_{ij}={i(N-j)\over N}~,\quad i \leq j~,
\ee
with the other components fixed by $A^{-1}_{ji}=A^{-1}_{ij}$.

Expanding a weight $\lambda$ in terms of the fundamental weights defines the Dynkin labels $d_i$,
\be
\lambda = \sum_i d_i \omega^{(i)}~.
\ee
The Dynkin labels associated with the weights of finite dimensional irreducible representations are integers.    A representation whose highest weight has Dynkin labels $(d_1, d_2, \ldots, d_{N-1})$ is associated with the Young tableau that has $d_1$ columns of height $1$, $d_2$ columns of height $2$,  and so on.

The Weyl vector is defined as
\be
\rho = \sum_i \omega^{(i)}~,
\ee
i.e. all its Dynkin labels equal $1$.  It obeys
\be
\rho \cdot \rho = \sum_{ij}A^{-1}_{ij}= {N(N^2-1)\over 12}~.
\ee
We also note
\be
\lambda \cdot \rho = {1\over 2} \sum_j j(N-j)\lambda_j~.
\ee

The quadratic Casimir for a representation with highest weight $\lambda$ is (up to choice of normalization)
\be
 C_2(\lambda) = \lambda \cdot (\lambda+2 \rho)~.
\ee
which can be evaluated using the above formulas.  In terms of generators in the Cartan-Weyl basis we have
\be
C_2 = \sum_i H^i H^i +\sum_{\alpha>0} (E^{\alpha}E^{-\alpha} + E^{-\alpha}E^{\alpha})~.
\ee

Besides the Cartan-Weyl basis, another basis is frequently used in the higher spin gravity context.  Namely, we start from an SL(2) subalgebra with generators $(L_1, L_0, L_{-1})$ obeying $[L_i,L_j]=(i-j)L_{i+j}$.    This subalgebra can be chosen so that the remaining generators lie in irreducible spin-$j$ representations of SL(2), with $j=3, 4, \ldots N$,  with one representation for each such $j$.   We  denote the SL(N) generators in this basis as $V^s_m$, with $s=2, 3, \ldots N$ and $m= -(s-1), -(s-2), \ldots, (s-2), (s-1)$.    Note that $V^2_i = L_i$.  We have
\be
[L_m, V^s_n]=(m(s-1)-n)V^s_{m+n}~.
\ee
The full SL(N) algebra is
\begin{equation}
\left[ V_{m}^s,V_{n}^{t}\right] =\sum\limits_{u=2,4,6,...}^{s+t-|s-t|-1}g_{u}^{st}\left(
m,n;N\right) V_{m+n}^{s+t-u} \text{.}
\end{equation}
The structure constants are denoted by $g_{u}^{st}\left(m,n;N\right)$, and can be written as the product
\begin{equation}
g_{u}^{st}\left( m,n;N\right) =\frac{q^{u-2}}{2\Gamma \left(
u\right) }\phi _{u}^{st}\left( N\right) {\cal N}_{u}^{st}\left(
m,n\right) \text{ ,}
\end{equation}
where
\begin{equation}\begin{aligned}\label{c3}
\phi_{u}^{st}\left( N\right)  =&\pFq{4}{3} {\frac{1}{2}-N,\frac{1}{2}+N,\frac{2-u}{2},\frac{1-u}{2}}{\frac{3}{2}-s,\frac{3}{2}-t,\frac{1}{2}+s+t-u}{1} \\
{\cal N}_{u}^{st}\left( m,n\right)  =&\sum\limits_{k=0}^{u-1}\left(
-1\right) ^{k}{{u-1}\choose k}
 \left( 1-s-m\right) _{u-1-k}\left( 1-s+m\right) _{k}
\left( 1-t-n\right) _{k}\left( 1-t+n\right)
_{u-1-k} \text{ ,}
\end{aligned}\end{equation}
$(a)_n=\Gamma(a+n)/\Gamma(a)$ is the rising Pochhammer symbol. $q$ is a normalization constant that can be scaled away by taking $V^s_m\rightarrow q^{s-2}V^s_m$. In the body of the paper, we take $q=1/4$, as in many other works (e.g. \cite{Gaberdiel:2011wb, Kraus:2011ds}).

The explicit form of these generators in the defining representation can be found in e.g. \cite{Campoleoni:2013lma}.  Here we just note that $L_0$ takes the form
\be\label{L0form}
L_0  = {\rm diag}\left({N-1\over 2}, {N-3\over 2}, \ldots, -{N-1\over 2}\right)~.
\ee
An important fact is that $L_0$ is the Cartan element dual to the Weyl vector,
\be
\rho \cdot H = L_0~.
\ee
This is the reason why the Weyl vector appears as the weight for the Wilson line probe that computes entanglement entropy, namely since  it should carry vanishing charge under $V^s_0$ for $s>2$. Note that $L_0 |\rho\rangle = \rho\cdot \rho |\rho\rangle = {N(N^2-1)\over 12}|\rho\rangle$.

\section{Deriving $P_0$}\label{P_0}
In this Appendix we derive the coefficients in the expansion \eqr{probe1}, which we rewrite here:
\e{e1}{P_0 = \sum_{s=2}^N {q_s\over \Tr_{\square}(V^s_0V^s_0)}V^s_0~.}
Recall that the $q_s$ are defined as the charges of a probe dual to a CFT operator $\Oc_L$: in particular, $q_s$ equals the zero mode charge under $V^s_0$, rescaled by $k_{CS}$ but without further normalization, as in \eqr{hlcharges}. These charges are fixed in the semiclassical limit \eqr{sclim}.

By definition, $P_0$ is a sum of Cartan elements, which we may parameterize as
\e{e2}{P_0 =  \sum_{s=2}^N {\alpha_s q_s}V^s_0~.}
Our goal is to compute the coefficients $\alpha_s$. As shown in \cite{Ammon:2013hba}, a defining property of $P_0$ is the relation $\Tr_{\square}(P_0P_0) = c_2$, where $c_2$ is the quadratic Casimir of SL(N) acting on the highest weight state $|\Oc_L\rangle$. From this we have
\e{e3}{c_2 = \sum_{s=2}^N \a_s^2 q_s^2 \Tr_{\square}(V^s_0V^s_0)~.}

On the other hand, consider the independent definition of the Casimir. Using notation $T_m \equiv \lbrace V^s_m \rbrace$, we have
\e{e4}{c_2 = g^{mn} T_mT_n~, \quad \text{where}\quad g_{mn} = \Tr_{\square}(T_mT_n)~.}
At large charges, its action on a highest weight state is
\e{e5}{c_2 \rar \sum_{s=2}^N {q_s^2\over  \Tr_{\square}(V^s_0V^s_0)}~ +~(\text{subleading})~.}
These are the contributions from $m=n=0$ terms; all subleading terms come from commutators. Equating \eqr{e3} and \eqr{e5}, we arrive at \eqr{e1}. One can check this against the $N=3$ results in Section 6 of \cite{deBoer:2014sna}.

\section{Details of SL(N) Wilson line computation}
\label{SLNappendix}

The result stated in section \ref{compG} relies on the computation of $\det X$, where $X$ is a $k\times k$ matrix with entries
\be
X_{ij}= \langle N+1-i | e^{\Lambda} |j\rangle~,\quad   i, j=1, 2, \ldots k~,
\ee
with
\be
\Lambda  = \left(L_1+ \sum_{s=2}^{N} Q_s V^s_{-(s-1)}
\right)w~.
\ee
The result will be expressed in terms of the eigenvalues of $\Lambda$ in the defining representation,
\be
{\rm eig}(-\Lambda) =  (\lambda_1, \lambda_2, \ldots \lambda_N)~.
\ee
What follows is a generalization of computations in \cite{Castro:2011iw,Hijano:2013fja}.

We begin by writing
\begin{align}
e^\Lambda = B K B^{-1}
\end{align}
where $B$ is an upper triangular matrix and $K$ can be diagonalized by a Vandermonde matrix. We then have
\begin{align}
X_{ij} =\sum_{lm} B_{N+1-i,l}K_{lm}B^{-1}_{mj}~.
\end{align}
Since B is upper triangular, the matrix elements are non zero only for $l \leq N+1-i$. As $i = 1,\ldots,k$, we have $l\geq N+1-k$. So the only part contributing to the determinant is a $k \times k$ block of the full matrix $B$ (explicitly $B_{i'l}$ with $i',l \geq N+1-k$). A similar analysis holds for $B^{-1}$, where now $m \leq k$ and the block that contributes is $B^{-1}_{mj'}$ with $m,j' \leq k$. Let the restriction of $B$ and $B^{-1}$ to their contributing blocks be $\widetilde{B}$ and $\widetilde{B^{\scriptscriptstyle -1}}$ respectively. The determinant then reduces to
\begin{align}\label{det}
\det X = \det(\widetilde{B}\widetilde{K}\widetilde{B^{\scriptscriptstyle -1}})
\end{align}
where now $K$ is automatically restricted to a $k\times k$ matrix $\widetilde{K}$. Hence, the determinant neatly factorizes into a product of $k\times k$ determinants.

$B$ obeys the following easily established properties \cite{Hijano:2013fja}
\begin{align}
B^{-1}_{jj} = \frac{1}{B_{jj}}, \hspace{1pc} B_{jj} = \sqrt{\frac{(N-1)!(j-1)!}{(N-j)!}} B_{11}~.
\end{align}
Since $\widetilde{B}$ and $\widetilde{B^{\scriptscriptstyle -1}}$ are also upper triangular their determinants depend only on their diagonal elements.  These are readily computed to give
\begin{align}\label{coefficient}
\det(\widetilde{B})\det(\widetilde{B^{\scriptscriptstyle -1}}) &= \prod_{j=1}^{k}B_{N+1-j,N+1-j}\prod_{i=1}^{k}B^{-1}_{ii}
= \prod_{j=1}^{k}\frac{\Gamma(N-j+1)}{\Gamma(j)}
\end{align}

We are then left with calculating $\det(\widetilde{K})$ in \eqref{det}, which is more involved.   Recall that the matrix $K$ is diagonalized by a Vandermonde matrix
\be
K = V e^{-\lambda} V^{-1}
\ee
where $\lambda = {\rm diag}(\lambda_1, \ldots, \lambda_N)$ and
\be
 V = \left(   \begin{array}{cccc}\lambda_1^{N-1} & \lambda_2^{N-1} & \ldots &\lambda_N^{N-1} \\
\vdots &\vdots & \ldots & \vdots\\
\lambda_1 & \lambda_2 & \ldots & \lambda_N \\
1 & 1& \ldots & 1\end{array}\right)~.
\ee
We therefore also have
\begin{align}\label{K}
\widetilde{K}_{lm} = \sum_{j=1}^{N} V_{lj}e^{-\lambda_j}V^{-1}_{jm}
\end{align}
where $l\geq N+1-k$ and $m \leq k$. Its determinant is
\begin{align}\label{vandermonde}
\det \widetilde{K} &= \frac{1}{k!} \epsilon_{a_1\ldots a_k} \epsilon_{b_1\ldots b_k} \widetilde{K}_{a_1 b_1}\ldots \widetilde{K}_{a_k b_k}\nonumber
\\
&= \frac{1}{k!}\sum_{j_1,\ldots, j_k} (\epsilon_{a_1\ldots a_k}V_{a_1 j_1}\ldots V_{a_k j_k})(\epsilon_{b_1\ldots b_k}V^{-1}_{j_1 b_1}\ldots V^{-1}_{j_k b_k}) e^{-\lambda_{j_1}}\ldots e^{-\lambda_{j_k}}~.
\end{align}
Note that $a_i \geq N+1-k$ restricts the matrix $V$ to the last $k$ rows and $b_i\leq k$ restricts $V^{-1}$ to the first $k$ columns. The first term in parenthesis, up to sign, is a determinant which we denote as $V[\lambda_{j_1},\ldots,\lambda_{j_k}]$,
\begin{align}\label{V def}
V[\lambda_{j_1},\ldots,\lambda_{j_k}] = \epsilon_{a_1\ldots a_k}V_{a_1 j_1}\ldots V_{a_k j_k} =
\begin{vmatrix}
\lambda^{k-1}_{j_1} & \ldots & \lambda^{k-1}_{j_k}\\
\vdots & \ddots & \vdots\\
\lambda_{j_1} & \ldots & \lambda_{j_k}\\
1 & \ldots & 1
\end{vmatrix}~.
\end{align}
To fix the sign, note that the ordering of rows in the first term in parenthesis in \eqref{vandermonde} is the exact reverse of the ordering above. Taking this into account we have
\begin{align}\label{reduced}
\det \widetilde{K} = \frac{(-1)^{k(k-1)/2}}{k!}\sum_{j_1,\ldots, j_k} V[\lambda_{j_1},\ldots,\lambda_{j_k}](\epsilon_{b_1\ldots b_k}V^{-1}_{j_1 b_1}\ldots V^{-1}_{j_k b_k}) e^{-\lambda_{j_1}}\ldots e^{-\lambda_{j_k}}~.
\end{align}
The inverse of the Vandermonde matrix is given by a well known expression:
\begin{align}
V^{-1}_{jk} = (-1)^{k-1}\frac{\sum^{1\leq m_1 < \ldots < m_{k-1}\leq N}_{m_1,\ldots, m_{k-1} \neq j}\lambda_{m_1}\ldots\lambda_{m_{k-1}}}{\prod_{m\neq j} (\lambda_{j}-\lambda_{m})}~,\quad m_i = 1,\ldots, N~.
\end{align}
Going back to  \eqref{reduced}, we write the term in parenthesis as
\begin{align}\label{inverse}
\widetilde{V^{\scriptscriptstyle -1}} =\epsilon_{b_1\ldots b_k}V^{-1}_{j_1 b_1}\ldots V^{-1}_{j_k b_k}= \begin{vmatrix}
V^{-1}_{j_11} & \ldots & V^{-1}_{j_1k}\\
\vdots & \ddots & \vdots \\
V^{-1}_{j_k1} & \ldots & V^{-1}_{j_kk}\\
\end{vmatrix}~.
\end{align}
We can pull out a factor of $\prod_{a_i\neq j_i}\frac{1}{\lambda_{j_i}-\lambda_{a_i}}$ from each row giving
\begin{align}\label{reduced inverse}
\widetilde{V^{\scriptscriptstyle -1}} = \begin{vmatrix}
V'_{j_11} & \ldots & V'_{j_1k}\\
\vdots & \ddots & \vdots \\
V'_{j_k1} & \ldots & V'_{j_kk}\\
\end{vmatrix} \left[\prod_{a_1\neq j_1}\frac{1}{\lambda_{j_1}-\lambda_{a_1}}\right]\ldots\left[\prod_{a_k\neq j_k}\frac{1}{\lambda_{j_k}-\lambda_{a_k}}\right]
\end{align}
where the $V'$ is defined by
\begin{align}\label{V' def}
V'_{jk} = (-1)^{k-1}\sum_{m_1,\ldots, m_{k-1} \neq j}\lambda_{m_1}\ldots\lambda_{m_{k-1}}~.
\end{align}
The sum in the above equation is over distinct $m_i$. To understand the structure of this determinant, consider the first few columns
\begin{align}
V'_{j1} = 1, \hspace{1pc} V'_{j2} = -(\sum_m \lambda_m - \lambda_j), \hspace{1pc} V'_{j3} = \sum^{1\leq m_1<m_2\leq N}_{m_1, m_2 \neq j}\lambda_{m_1}\lambda_{m_2}~.
\end{align}
$V'_{j3}$ is the sum of all possible pairs of distinct eigenvalues excluding all pairs with $\lambda_j$. Generalizing, $V'_{jk}$ is the sum of all possible groups of $(k-1)$ eigenvalues excluding all groups with  $\lambda_j$ in them. This can be written concisely as
\begin{align}\label{elements}
V'_{jk} = (-1)^{k-1}\left(\sum \lambda_{m_1}\ldots\lambda_{m_{k-1}}\right) + \lambda_j V'_{j,k-1}
\end{align}
where the sum is over all distinct $m_i$ and the second term gives all groups of size $(k-1)$ containing $\lambda_j$.

We are now ready to evaluate the determinant in \eqref{reduced inverse}. We will see that it reduces to the Vandermonde determinant defined in \eqref{V def}. Perform the following row and column operations on the determinant
\begin{subequations}
\begin{align}
C_m &\rightarrow C_m - \lambda_1 C_{m-1}\label{1}\\
R_n &\rightarrow R_n - R_1\label{2}\\
C_m &\rightarrow C_m - \alpha_m C_1\label{3}~.
\end{align}
\end{subequations}
The first column operation \eqref{1} changes $V'_{jk} \rightarrow (-1)^{k-1}\left(\sum \lambda_{m_1}\ldots\lambda_{m_{k-1}}\right) + (\lambda_j-\lambda_1) V'_{j,k-1}$. Note that the first row only has the sum over all possible groups of size $(k-1)$. The row operation \eqref{2} then subtracts this sum from all the other rows giving $V'_{jk}\rightarrow (\lambda_j-\lambda_1) V'_{j,k-1}$. It also kills all but the first element of the first column. The determinant then looks like
\begin{align}
\begin{vmatrix}
1 & -\sum_m \lambda_m & \sum_{m_1<m_2}\lambda_{m_1}\lambda_{m_2} & \ldots\\
0 & (\lambda_2-\lambda_1) V'_{21} & (\lambda_2-\lambda_1) V'_{22} & \ldots\\
0 & (\lambda_3-\lambda_1) V'_{31} & (\lambda_3-\lambda_1) V'_{32} & \ldots\\
\vdots & \vdots & \vdots & \ddots\\
0 & (\lambda_k-\lambda_1) V'_{k1} & (\lambda_k-\lambda_1) V'_{k2} & \ldots\\
\end{vmatrix}~.
\end{align}
We then choose $\alpha_m$ in \eqref{3} to kill all but the first term in the first row. Factoring out $\prod_{m\neq 1}(\lambda_m-\lambda_1)$, the determinant reduces to that of a $(k-1)\times(k-1)$ matrix of the same kind as before which is independent of $\lambda_1$. Continuing all the way to the $1\times 1$ case, we then have (from \eqref{V def}, \eqref{reduced inverse} and \eqref{V' def})
\begin{align}\label{inverse result}
\widetilde{V^{\scriptscriptstyle -1}} = V[\lambda_{j_1},\ldots,\lambda_{j_k}] \left[\prod_{a_1\neq j_1}\frac{1}{\lambda_{j_1}-\lambda_{a_1}}\right]\ldots\left[\prod_{a_k\neq j_k}\frac{1}{\lambda_{j_k}-\lambda_{a_k}}\right]~.
\end{align}
Combining everything (\eqref{coefficient}, \eqref{reduced}, \eqref{inverse result}), we find
\begin{align}
\begin{split}
\det X = \frac{(-1)^{k(k-1)/2}}{k!}\left[\prod_{j=1}^{k}\frac{\Gamma(N-j+1)}{\Gamma(j)}\right]\sum_{j_1,\ldots, j_k} V[\lambda_{j_1},\ldots,\lambda_{j_k}]^2 \left[\prod_{a_1\neq j_1}\frac{1}{\lambda_{j_1}-\lambda_{a_1}}\right]\\\ldots\left[\prod_{a_k\neq j_k}\frac{1}{\lambda_{j_k}-\lambda_{a_k}}\right]e^{-\lambda_{j_1}}\ldots e^{-\lambda_{j_k}}
\end{split}~.
\end{align}
Absorbing the $k!$ into the product,
\begin{align}
\begin{split}
\det X = (-1)^{k(k-1)/2}\left[\prod_{j=1}^{k}\frac{\Gamma(N-j+1)}{\Gamma(j+1)}\right]\sum_{j_1,\ldots, j_k} V[\lambda_{j_1},\ldots,\lambda_{j_k}]^2 \left[\prod_{a_1\neq j_1}\frac{1}{\lambda_{j_1}-\lambda_{a_1}}\right]\\\ldots\left[\prod_{a_k\neq j_k}\frac{1}{\lambda_{j_k}-\lambda_{a_k}}\right]e^{-\lambda_{j_1}}\ldots e^{-\lambda_{j_k}}
\end{split}
\end{align}
which is the result appearing in (\ref{Gkresult}).

\section{Spin-$s$ charge in an antisymmetric tensor representation}\label{probe charge}
The goal of this section is to derive the charge carried by the probe Wilson line in the ${\bf {\rm \bf asym}_k}$ representation, as given in Section \eqr{probe charges}. To calculate this, consider the chiral deformation connection $a_w = V^2_1, a_{\overline{w}} = -\mu V^s_{-(s-1)}$. The Wilson line in an arbitrary representation is given by
\begin{align}
\label{apa}
\Gc_{R} = \braket{-\text{hw}_R | e^{w V^2_1 - \mu \overline{w} V^s_{s-1}} | \text{hw}_R}
\end{align}
where $\ket{\text{hw}_R}$ is the highest weight state of the representation $R$.  The term linear in $\mu$ in the Wilson line is proportional to the spin-$s$ charge carried by the Wilson line.  Further, it is sufficient to focus on the antisymmetric tensor representations: an arbitrary representation can be written as tensor products of antisymmetric tensor representations and the Wilson line is given by
\begin{align}
\label{apb}
\Gc_{R} = \prod_{k=1}^{N-1} \Gc_k^{\,d_k}
\end{align}
where $d_k$ are the Dynkin labels of the representation $R$. For the {\bf asym}$_{\bf k}$ representation, $d_i=\delta_{i,k}$, the Young tableau is a single column with $k$ boxes and the highest weight state can be written in terms of states in the defining representation
\begin{align}
\label{apc}
\begin{split}
\ket{\text{hw}} = \frac{1}{\sqrt{k!}}\epsilon_{i_1\ldots i_k} \ket{i_1}\ldots\ket{i_k},\hspace{0.5pc}
\ket{-\text{hw}} = \frac{1}{\sqrt{k!}}\epsilon_{i_1\ldots i_k} \ket{N-i_1+1}\ldots\ket{N-i_k+1}
\end{split}
\end{align}
where $\ket{1}$ and $\ket{N}$ are the highest and lowest weight states of the defining representation respectively. As seen in section \ref{closed},  the Wilson line in an antisymmetric tensor representation is equal to the determinant of a matrix formed from matrix elements in the defining representation,
\begin{align}
\label{apd}
\Gc_k = \det\tilde{A}~,\quad  \tilde{A}_{i,j}=\braket{N-i+1 | e^{w V^2_1} e^{-\mu \overline{w} V^s_{s-1}} | j}~,\quad i,j =1, \ldots, k
\end{align}
Further, in the defining representation we have $V^s_{s-1} = (V^2_1)^{s-1}$, greatly simplifying the calculation. Using this, the matrix elements to first order in $\mu$ are found to be
\begin{align}
\label{ape}
\tilde{A}_{i,j} = w^{N+1-i-j} \sqrt{\frac{(N-j)!}{(j-1)!}\frac{(N-i)!}{(i-1)!}}\left(\frac{1}{(N+1-i-j)!}-\frac{1}{(N-(s-2)-i-j)!}\frac{\mu \overline{w}}{w^{s-1}}
\right)
\end{align}
It is instructive to first compute the zeroth order piece. It will turn out that a very similar procedure will also be applicable to the term linear in $\mu$. First note that some of the terms depend on $i$ and $j$ separately. We can factor these out from each row and column of the determinant and define a new matrix element
\begin{align}
\label{apea}
A_{i,j} = \frac{(N-j)!}{(N+1-i-j)!}
\end{align}
where a factor of $(N-j)!$ has been added for convenience. The two determinants are related by
\begin{align}
\label{apeb}
\det \tilde{A} = w^{k(N-k)}\prod_{i=1}^k \frac{1}{(i-1)!}~\det A
\end{align}
Performing the column operation $C_j \to C_j - C_{j-1}$ in $\det A$ reduces the $k\times k$ determinant to a $(k-1) \times (k-1)$ determinant of the same type with $N\to N-1$. We can then continue performing column operations all the way up to the $1\times 1$ case which is trivial.
\begin{align}
\label{apec}
\det_{k\times k} A [N] = (-1)^{k-1} (k-1)! \det_{k-1\times k-1} A[N-1]
\end{align}
Putting all of this together, we find
\begin{align}
\label{aped}
\Gc_k = w^{k(N-k)}(-1)^{k(k-1)/2} + \mathcal{O}(\mu)
\end{align}

For the term linear in $\mu$, perform the same factorization and define a reduced matrix element by
\begin{align}
\label{apf}
A_{i,j} = \frac{(N-j)!}{(N+1-i-j)!} - \frac{(N-j)!}{(N-(s-2)-i-j)!}\frac{\mu \overline{w}}{w^{s-1}} + \mathcal{O}(\mu^2)
\end{align}
As before the two determinants are related by
\begin{align}
\label{apg}
\det \tilde{A} = w^{k(N-k)} \prod_{i=1}^k \frac{1}{(i-1)!}~\det A
\end{align}
Let's denote the terms in \eqref{apf} as $A^0_{i,j}$ and $A^1_{i,j}$, with $A^0$ being the term independent of $\mu$. The term in the Wilson line independent of $\mu$ is calculated by considering the determinant with all matrix elements being independent of $\mu$, i.e. $A^0_{i,j}$. Note that the determinant is linear in its rows. Hence, the term linear in $\mu$ is given by determinants where one of the rows in the $\mu$ independent determinant is replaced by terms linear in $\mu$, i.e. $A^1_{i,j}$. Further, the matrix elements $A^1$ are of the same form as $A^0$ except with $i \to i + (s-1)$. In other words, the replaced row is proportional to the $(s-1)^{\text{th}}$ row below it. All the first order determinants must then vanish except when $i+(s-1) > k$. So, the whole calculation boils down to calculating $(s-1)$ such determinants.

Let the replaced row be the $(k-r)^{\text{th}}$ row ($0 \leq r \leq s-2$) and the matrix be denoted by $A_r$. Considering the matrix elements in \eqref{apf}, we can perform the column operations $C_j \to C_j - C_{j-1}$ which reduces the determinant effectively to that of a $(k-1)\times (k-1)$ matrix. Each column operation also effectively reduces the values of $N$, $i$ and $j$. This proceeds as in the $\mu$ independent calculation except for a factor out front. We then perform the column operation repeatedly until we hit the replaced row,
\begin{align}
\label{aph}
\begin{split}
\det_{k\times k} A_r[N] &= (-1)^{k-1} (k-1)! \frac{k-r+s-2}{k-1-r} \det_{k-1\times k-1} A_r[N-1]
\\
&= \left(\prod_{i=r+2}^k(-1)^{i-1} (i-1)!\right)~\frac{(k-r+s-2)!}{(k-1-r)!(s-1)!}~\det_{r+1\times r+1}A_r[N-k+r+1]
\end{split}
\end{align}
To calculate the $(r+1)\times(r+1)$ determinant, we first exchange rows such that the replaced row is at the bottom and then factor out $(N-k+r+1-j)$ from each column. To elaborate, the matrix elements (ignoring the last row) before factoring out looked like $(N'+1-j)!/(N'+2-i-j)!$ where $N' = N-k+r$ and $i$ starts from 2 due to the row exchanges performed earlier. By factoring out $(N'+1-j)$, the matrix elements are of the same form as the $\mu$ independent matrix elements with $N \to N'$ and with $i$ now starting from 1.
\begin{align}
\label{api}
\det_{r+1\times r+1} A_r[N-k+r+1] = (-1)^r \prod_{j=1}^{r+1}(N-k+r+1-j) ~\det_{r+1\times r+1} A'[N-k+r]
\end{align}
where the sign on the right hand side is from exchanging rows. We can then perform the same column operations as before all the way until the last row.
\begin{align}
\label{apj}
\begin{split}
\det_{r+1\times r+1} A'[N-k+r] &= (-1)^r r! \frac{(s-2)}{r} \det_{r\times r} A'[N-k+r-1]
\\
&= \left(\prod_{i=2}^{r+1} (-1)^{i-1}(i-1)!\right)\frac{(s-2)!}{(s-2-r)!r!} \det_{1\times 1}A'[N-k]
\end{split}
\end{align}
We can now put all of it together from \cref{aph,api,apj} to get
\begin{align}
\label{apk}
\det_{k\times k} A[N] = w^{k(N-k)}(-1)^{k(k-1)/2} \frac{(-1)^r}{(s-1)} \frac{(k-r+s-2)!}{(k-1-r)!}\frac{1}{(s-2-r)!r!}\frac{N'!}{(N'-s+1)!}
\end{align}
where $N' = N-k+r$. The full term linear in $\mu$ is given by adding up $(s-1)$ such determinants ($r = 0,\ldots, s-2$). Up to $N$ and $k$ independent proportionality constants, the charge carried by the probe Wilson line in an {\bf asym}$_{\bf k}$ representation is given by
\begin{align}
\label{apl}
q_{s,k} \sim \sum_{r=0}^{s-2} \frac{(-1)^r}{(s-1)} \frac{(k-r+s-2)!}{(k-1-r)!}\frac{1}{(s-2-r)!r!}\frac{N'!}{(N'-s+1)!}
\end{align}
where it is to be understood that $1/n! = 0~ \forall~ n \in \mathbb{Z}^-$.

This formula for the charge is not very illuminating. To improve this, we note that a connection valued in sl(N) consists of higher spin fields with spins $2\leq s \leq N$ only. A simple check of our result is that the higher spin charge must vanish if we choose $N < s$. To make this feature manifest we rewrite \eqref{apl} as
\begin{align}
\label{gzn2}
q_{s,k} \propto \frac{\Gamma(N-k+1)}{\Gamma(N-s+1)}\sum_{r=0}^{s-2}(-1)^r\frac{(s-1-r)_k(N-k+1)_r(N-s+2-k+r)_{k-1-r}}{(s-1)\Gamma(k-r)\Gamma(r+1)}
\end{align}
where $(a)_n = a(a+1)\ldots (a+n-1)$ is the ascending Pochhammer symbol. Equation \eqr{gzn2} was presented earlier as equation \eqr{gzn}.  All the $N$ dependence inside the sum is polynomial. It is then clear that the higher spin charges vanish for $k<N<s$.

The continuation to Vasiliev, $N\rightarrow-\lambda$, is straightforward and we obtain \eqref{gzna}. The results for some low-lying values of $s$ and $k$ are presented below:
\begin{align}
\label{gzo}
\begin{split}
q_{s,1} &= \frac{\Gamma(s)^2}{\Gamma(2s-1)}\frac{\Gamma(s+\lambda)}{\Gamma(1+\lambda)}
\\
q_{s,2} &= \frac{\Gamma(s)^2}{\Gamma(2s-1)}\frac{\Gamma(s+\lambda)}{\Gamma(2+\lambda)}(2\lambda+s^2-s+2)
\\
q_{3,k} &= \frac{1}{6} k (\lambda+k)(\lambda+2k)
\\
q_{4,k} &=\frac{1}{20} k (\lambda+k)(\lambda^2+5k\lambda+5k^2+1)
\\
q_{5,k} &= \frac{1}{70} k(\lambda+k)(\lambda+2k)(\lambda^2+7k\lambda+7k^2+5)
\end{split}
\end{align}

\section{Prescriptions for defining and computing entanglement and thermal entropies}
\label{prescrips}

On both the bulk and CFT sides there is a number of distinct prescriptions that one adopts in the computation of thermal and entanglement entropies.  This can lead to confusion when comparing results.  In the case of thermal entropy this is now fairly well understood, but open questions remain regarding entanglement entropies. In this paper we have taken something of an experimental approach when comparing results, noting which prescriptions seem to match up with each other.  We will not give a full accounting here of the possibilities, but just mention the basic issues that arise.

\vspace{.2cm}
\noindent
{\underline{\bf Holomorphic versus canonical}}
\vspace{.2cm}

This issue concerns the gauge freedom in writing Chern-Simons connections, and the map between bulk and CFT charges and potentials.  The holomorphic prescription was originally proposed in \cite{Gutperle:2011kf}.  Here one puts $a_w$ in highest weight gauge, with the charges appearing as the coefficients of the generators with negative mode index.  The potentials then appear in $a_{\wb}$. This is justified by the fact that the flatness conditions then match with CFT Ward identities computed from the path integral.  A formula for the black entropy was proposed in \cite{Gutperle:2011kf} and found to agree with the thermal CFT entropy in \cite{Kraus:2011ds,Gaberdiel:2012yb}.  The holomorphic black hole entropy formula reads
\bea\label{Shol}
S_{\rm hol} = -2\pi i k_{CS} \Tr[h a_w] +2\pi i k_{CS}\Tr [\overline{h}\overline{a}_{\wb}]~.
\eea
Here $h$ and $\overline{h}$ denote holonomies around the thermal circle of the Euclidean black hole.
However, it was pointed out \cite{Perez:2013xi,deBoer:2013gz} that canonical methods for computing black hole entropy yield not (\ref{Shol}) but rather this formula but with $a_w$ and $\overline{a}_{\wb}$ replaced by $a_\phi$ and $\overline{a}_\phi$.  Call this formula $S_{\rm can}$.  Since $S_{\rm hol}\neq S_{\rm can}$, this would seem to favor $S_{\rm hol}$ based on its agreement with the CFT.  However, it was eventually realized that is natural to reexpress $S_{\rm can}$ in terms of a new set of charges obtained by gauge transforming away $a_{\wb}$ and $\ab_w$, and that the resulting expression has the same functional form as $S_{\rm hol}$.  We refer to \cite{Compere:2013nba,Hijano:2014sqa} for more discussion of this.

The same basic issue arises in the definition of Wilson lines and  entanglement entropy. The Wilson line which, for a spacelike geodesic, involves integrating $a_\phi$, is called canonical, but as noted in \cite{deBoer:2013vca} one can define a holomorphic Wilson line by integrating $a_w$.   In the large interval limits, these Wilson lines compute the canonical and holomorphic entropies, and then they are related as noted above.     However, for a finite interval there is less understanding about the relation of the two prescriptions.  In particular, in the presence of higher spin potentials the entanglement entropy computed in the holomorphic prescription has been found to match against CFT results \cite{Datta:2014ska,Datta:2014uxa}, but the same is not true for the canonical prescription.

\vspace{.2cm}
\noindent
{\underline{\bf Hamiltonian trace versus Lagrangian path integral}}
\vspace{.2cm}

Another issue to be aware of is the nontrivial relationship between a path integral for a CFT deformed by potentials for higher spin charge and a canonical trace formula.  For standard field theories with two derivative actions there is of course a simple relationship between the two, but higher spin potentials introduce higher derivatives in the action.  A detailed discussion of the issues involved appears in \cite{deBoer:2014fra}.

\vspace{.2cm}
\noindent
{\underline{\bf Integration prescriptions}}
\vspace{.2cm}

If one performs conformal perturbation theory in the CFT with respect to potentials for higher spin currents then one has to perform integrals of current correlators.  Such integrals are divergent due to short distance singularities, and an integration prescription is required to make sense of them.  At finite temperature the CFT is defined on the torus, and several possibilities are available \cite{Datta:2014ska,Datta:2014uxa,Datta:2014zpa}.  In this paper we perform perturbation theory at zero temperature corresponding to the plane and employ a different prescription, as discussed in Section \ref{CFTmatch}.

\section{Calculational details of Section \ref{vasiliev}}
\subsection{Small charge expansion}\label{smallchargedetails}

In section \ref{addinghsQ} we defined quantities ${\cal J}_n$ as matrix elements needed in the computation of the small charge expansion.
Here we express these objects in a more convenient form. First we use SL(N) group theory to rewrite ${\cal J}_n$ in terms of the conjugated generators. Then, we use the properties of the highest and lowest weight states to recast the matrix element between the highest and lowest state into one between highest weight states. 

From \eqref{ep}, the ${\cal J}_n$ are given by
\begin{align}
\label{epa}
{\cal J}_n(x_i) \equiv \braket{-\rho | M_0(x_1) V^s_{-s+1} M_0(x_2)\ldots V^s_{-s+1} M_0(x_{n+1}) | \rho}
\end{align}
As noted before the $x_i$ are simply $n$ partitions of unity. We can exploit this to write \eqref{epa} as
\begin{align}
\label{eq}
\begin{split}
{\cal J}_n(x_i) =& \bra{-\rho}M_0(1)~M_0(-1+x_1)V^s_{-s+1}M_0(1-x_1)\\ &\ldots M_0(-1+\Sigma^n x_i)V^s_{-s+1} M_0(1-\Sigma^n x_i)\ket{\rho}
\end{split}
\end{align}
The $M_0$ are SL(2) elements and \{$V^s_m$\} for fixed $s$ are in irreducible representations of SL(2), in particular the spin $s-1$ representation. The transformation of these elements are easily obtained by noting that the transformed elements satisfy first order differential equations i.e.
\begin{align}
\label{er}
\frac{d}{dt}V^s_m(t) = [\Lambda_0, V^s_m(t)]~,~~~V^s_m(t) := e^{t\Lambda_0} V^s_me^{-t\Lambda_0}
\end{align}
Writing the exponential of an sl(2) element as product of exponentials,\footnote{It is now more convenient to write $e^{\Lambda_0} = e^{\tilde{c}_{-1} V^2_{-1}} e^{\log \tilde{c}_0 V^2_0} e^{\tilde{c}_1 V^2_1}$ where now $\tilde{c}_0 = c_0^{-1}, \tilde{c}_1 = c_1$} the object in \eqref{eq} becomes
\begin{align}
\label{es}
\begin{split}
{\cal J}_n(x_i) = c_0^{h_\rho}\braket{-\rho | e^{c_1 V^2_1}~V^s_{-s+1}(-1+x_1)\ldots V^s_{-s+1}(-1+\Sigma^n x_i) | \rho}
\end{split}
\end{align}
The highest weight state $\ket{\rho}$ and the lowest weight state $\ket{-\rho}$ are related by repeated action of $V^2_{\pm 1}$.  Using
\be
\langle -\rho|(V^2_1)^{2h_\rho}|\rho\rangle=(2h_\rho)!~,\quad \langle\rho | (V^2_{-1})^n (V^2_1)^n | \rho\rangle =  \frac{(-1)^n(2h_\rho)!n! }{(2h_\rho-n)!}
\ee
it is straightforward to verify
\be\label{rhorel}
\bra{-\rho} e^{c_1 V^2_1}= c_1^{2h_\rho}\bra{\rho} e^{-\tfrac{1}{c_1} V^2_{-1}}
\ee
We can then rewrite the matrix element in \eqref{es} as
\begin{align}
\label{ew2}
\begin{split}
{\cal J}_n(x_i) &= (c_0 c_1^2)^h \braket{\rho | e^{-\tfrac{1}{c_1} V^2_{-1}}~V^s_{-s+1}(-1+x_1)\ldots V^s_{-s+1}(-1+\Sigma^n x_i) | \rho}
\\
&= \braket{\rho | \tilde{V}^s_{-s+1}(-1+x_1)\ldots \tilde{V}^s_{-s+1}(-1+\Sigma_1^n x_i) | \rho}\Gc_\rho\big|_{Q_s=0}
\end{split}
\end{align}
where
\be
\tilde{V}(t) = e^{-1/c_1 V^2_{-1}}V(t)e^{1/c_1 V^2_{-1}} \ee
and we have used $V^2_{-1}\ket{\rho} = 0$ to
introduce an exponential on the right. Since $(V^2_{-1})^{2h_\rho+1}=0$, $\tilde{V}(t)$ is easily computed. The matrix elements can then be obtained by normal ordering the generators and using properties of the Weyl representation.

\subsection{Direct calculation in short interval expansion}\label{shortinterval}

Here we calculate \eqr{f3} directly, starting from \eqr{ezf}. We want to write $e^{w(V^2_1 + Q_3 V^3_{-2})}$ as a product of exponentials. For this we use the Zassenhaus formula, i.e. the ``reverse-BCH'' formula: given a function $e^{w(X+Y)}$ for $[X,Y]\neq 0$, one can write it as a generically infinite product
\e{she}{e^{w(X+Y)} = e^{wX} e^{wY} \prod_{n=2}^{\infty}e^{w^n c_n(X,Y)}}
where the $c_n(X,Y)$ are terms built out of $n-1$ nested commutators of $X$ and $Y$.  For example,
\e{shf}{c_2 = -{1\over 2}[X,Y]~, ~~ c_3= {1\over 3}[Y,[X,Y]] + {1\over 6} [X,[X,Y]]}
In perturbation theory in $Q_3$ (or $w$), the above product truncates at finite order due to the ad nilpotency of the SL(N) generators (cf. \eqr{adxy}.) . For the spin-3 perturbation considered here, the relevant fact is
\e{}{\text{ad}_{V^2_1}^{5}V^3_{-2}=0~.}
The $O(Q_3)$ term in \eqr{ezf} vanishes identically, due to \eqr{rhorho}. We focus on the $O(Q_3^2)$ term. At this order, the product in \eqr{she} truncates at $n=9$: at $n=10$, the $O(Q_3^2)$ term would have eight $V^2_1$'s and two $V^3_{-2}$'s, with which one cannot make a nonzero commutator. There exist efficient algorithms \cite{Casas} to find the functions $c_n(X,Y)$ and they can be evaluated using Mathematica. Using SL(N) commutators and the explicit formulae for the $c_n$ as given in \cite{Casas}, we find the following:
\es{shg2}{c_2 &= -2 Q_3 V^3_{-1}\\
c_3 &= 2 Q_3 V^3_0 - {8Q_3^2\over 3}V^4_{-3}\\
c_4 &= -Q_3 V^3_1 + 6Q_3^2 V^4_{-2}\\
c_5&={Q_3\over 5} V^3_2 + {Q_3^2\over 25}(-240 V^4_{-1}-2(N^2-4) V^2_{-1})\\
c_6&= {20Q_3^2\over 3} V^4_0+O(Q_3^3)\\
c_7 &= -{2Q_3^2\over 35}(70 V^4_1+(N^2-4)V^2_1)+O(Q_3^3)\\
c_8&=Q_3^2 V^4_2+O(Q_3^3)\\
c_9&=-{2Q_3^2\over 9}V^4_3+O(Q_3^3)}
These are now plugged into
\e{zass}{\Gc_{\rho}(Q_3) = \bra{-\rho}e^{w V^2_1} e^{w Q_3 V^3_{-2}}\prod_{n=2}^{\infty}e^{w^nc_n}\ket{\rho}}
Using \eqr{hwlw} and \eqr{ea}, we see that certain terms will annihilate $\ket{\rho}$ inside \eqr{zass}. Furthermore, any terms involving only a single $V^3_m$ or $V^4_m$ generator will also give zero due to \eqr{rhorho}. (This is the same reason the $O(Q_3)$ term vanishes.) So we can work with the following set instead:
\es{shi}{c_2 &= -2 Q_3 V^3_{-1}\\
c_3 &= 2 Q_3 V^3_0 \\
c_4 &= -Q_3 V^3_1\\
c_5&={Q_3\over 5} V^3_2 \\
c_7 &= -{2Q_3^2\over 35}(N^2-4)V^2_1}
All others vanish to this order inside \eqr{zass}. Using \eqr{ek} and the following commutators,
\es{shk}{V^3_{-2}V^3_2\rhor &= {4\over 5}(N^2-4)h_{\rho} \rhor\\
V^3_{-1}V^3_1\rhor &= -{N^2-4\over 10}h_{\rho} \rhor\\
V^3_{-1}V^3_2\rhor &= {N^2-4\over 5}V^2_1\rhor\\
V^3_0V^3_1\rhor &= -{N^2-4\over 10}V^2_1\rhor}
one arrives at
\e{}{\Gc_{\rho}(Q_3) = \Gc_{\rho}(0)\left(1-h_{\rho}{N^2-4\over 3500}Q_3^2 w^6+ O(w^{12})\right)}
Equation \eqr{f3} follows.

\subsec{Deriving Virasoro blocks}\label{wnvir}
Here we show how to get from \eqr{ezze2} to \eqr{Aform}.

To evaluate ${\cal J}_1$, we first define
\begin{align}
\label{ezzc}
f(\beta) \equiv \braket{1|e^{\beta \tilde{V}^2_{-1}(t)}|1}~.
\end{align}
To compute $f(\beta)$ we can follow the same procedure of rewriting the exponential as a product of exponentials.
\begin{align}
\label{ezzd}
\begin{split}
f(\beta) &= \braket{1 | e^{d_1 V^2_1} e^{\log d_0 V^2_0} e^{d_{-1} V^2_{-1}} | 1}
\\
&= d_0^{(N-1)/2}
\\
&= \left(1 + \tilde{\gamma} \frac{\beta}{\alpha}\right)^{N-1}~,\quad \text{where} \quad \tilde{\gamma} = 2 \csc \tfrac{w\alpha}{2}\sin \tfrac{w\alpha t}{2}\sin \tfrac{w\alpha(t+1)}{2}~.
\end{split}
\end{align}
Then
\begin{align}
\label{ezze}
\begin{split}
A_s(w) &= w \int^1_0 dt~\frac{\partial^{s-1}}{\partial \beta^{s-1}}f(\beta)\bigg|_{\beta = 0}
\\
&= \frac{w}{\alpha^{s-1}} \frac{(N-1)!}{(N-s)!} \int^1_0 dt~ \gamma^{s-1}
\end{split}
\end{align}
where $\gamma(t) = \tilde{\gamma}(t-1) = 2 \csc \tfrac{w\alpha}{2}\sin \tfrac{w\alpha t}{2}\sin \tfrac{w\alpha(t-1)}{2}$.

To derive a recursion relation for the integral in \eqref{ezze}, we first integrate by parts and then rewrite the new expression in terms of the old. Explicitly,
\begin{align}
\label{ezzh}
\begin{split}
I_s \equiv \int^1_0dt~\gamma^{s-1} &= \int^1_0dt~\gamma^{s-2}\frac{d}{dt}\left(\int\gamma\right)
\\&= -(s-2)\int^1_0dt~\gamma^{s-3}\left(\frac{d\gamma}{dt}\int\gamma\right)
\end{split}
\end{align}
where the boundary term vanishes as $\gamma(0)=\gamma(1)=0$ and $\int\gamma$ is the antiderivative of $\gamma$. Using the explicit form of $\gamma$ we find
\begin{align}
\label{ezzi}
\frac{d\gamma}{dt}\int\gamma = -\left(1+2\gamma \cot \frac{w\alpha}{2} - \gamma^2 -t\frac{d\gamma}{dt}\cot \frac{w\alpha}{2}\right)~.
\end{align}
The recursion relation then follows since all the $t$ dependence can be rewritten in terms of $\gamma$ and $t\frac{d\gamma}{dt}$. The integral in \eqref{ezzh} then becomes
\begin{align}
\label{ezzj}
(s-1) I_s = (2s-3) \cot \frac{w\alpha}{2}I_{s-1} + (s-2) I_{s-2}~.
\end{align}
It only takes a little bit of algebra to then show that these are almost the right recursion relations for $A_s$ of the form expected in (\ref{Aform}). Fixing an overall $s$-dependent factor using the recursion relation above and explicitly evaluating it for $s=2$ gives
\begin{align}
\label{ezzk}
A_s(w) = -\frac{\Gamma(s)\Gamma(s)}{\Gamma(2s)}\frac{\Gamma(N)}{\Gamma(N-s+1)}(i\alpha)^{-s}\left(1-e^{i\alpha w}\right)^s F(s,s;2s;1-e^{i\alpha w})~.
\end{align}
This indeed satisfies \eqr{Aform}. Note that the poles of the gamma function in the above formula enforce the condition $A_s=0$ for $s>N$ as we expect. 

\bibliographystyle{ssg}
\bibliography{biblio}

\end{document}